\documentclass[twocolumn]{aastex63}

\usepackage{amsmath}
\usepackage{txfonts}
\usepackage{xcolor}
\usepackage{graphicx}
\usepackage{gensymb}
\usepackage{booktabs}
\usepackage{lipsum}
\usepackage{bm}
\usepackage{physics}

\AtBeginDocument{\mathcode`v=\varv}

\newcommand\ab{\bm{a}}

\newcommand\rb{\bm{r}}

\newcommand\mub{\bm{\mu}}
\newcommand\sigmab{\bm{\sigma}}
\newcommand\thetab{\bm{\theta}}
\newcommand{\HI}{\ifmmode \mathrm{\ion{H}{1}} \else \ion{H}{1} \fi}
\newcommand{\HII}{\ifmmode \mathrm{\ion{H}{2}} \else \ion{H}{2} \fi}
\newcommand{\nh}{\ifmmode N_{{\mathrm{H}} \, \mathrm{I}} \else $N_{{\mathrm{H}} \, \mathrm{I}}$\fi} 
\newcommand{\mh}{\ifmmode M_{{\mathrm{H}} \, \mathrm{I}} \else $N_{{\mathrm{H}} \, \mathrm{I}}$\fi}

\def\GHz{\ifmmode $\,GHz$\else \,GHz\fi}
\def\MJysr{\ifmmode \,$MJy\,sr\mo$\else \,MJy\,sr\mo\fi}
\def\microns{\ifmmode \,\mu$m$\else \,$\mu$m\fi}

\def\kms{\ifmmode $\,km\,s$^{-1}\else \,km\,s$^{-1}$\fi}

\defcitealias{wolfire_neutral_2003}{W03}
\defcitealias{wolfire_neutral_1995}{W95a}
\defcitealias{wolfire_multiphase_1995}{W95b}
\defcitealias{marchal_rohsa:_2019}{M19}

\received{November 10, 2020}
\accepted{December 5, 2020}
\submitjournal{ApJ}

\shorttitle{Thermal and turbulent properties of the WNM}
\shortauthors{Marchal \& Miville-Desch\^enes}
\graphicspath{{./}{figures/}}

\begin{document}

\title{Thermal and turbulent properties of the Warm Neutral Medium in the solar neighborhood}

\correspondingauthor{Antoine Marchal}
\email{amarchal@cita.utoronto.ca}

\author[0000-0002-5501-232X]{Antoine Marchal}
\affiliation{Canadian Institute for Theoretical Astrophysics, University of Toronto, 60 St. George Street, Toronto, ON M5S 3H8, Canada}

\author[0000-0002-7351-6062]{Marc-Antoine Miville-Desch\^enes}
\affiliation{AIM, CEA, CNRS, Universit\'e Paris-Saclay, Universit\'e Paris Diderot, Sorbonne Paris Cit\'e, F-91191 Gif-sur-Yvette, France}

\begin{abstract}
The transition from the diffuse warm neutral medium (WNM) to the dense cold neutral medium (CNM) is what set the initial conditions to the formation of molecular clouds. The properties of the turbulent cascade in the WNM, essential to describe this radiative condensation process, have remained elusive in part due to the difficulty to map out the structure and kinematics of each \HI\ thermal phases. Here we present an analysis of a 
21\,cm hyper-spectral data cube from the GHIGLS \HI\ survey where the contribution of the WNM is extracted using {\tt ROHSA}, a Gaussian decomposition tool that includes spatial regularization. The distance and volume of the WNM emission is estimated using 3D dust extinction map information. The thermal and turbulent contributions to the Doppler line width of the WNM were disentangled using two techniques, one based on the statistical properties of the column density and centroid velocity fields, and another on the relative motions of CNM structures as a probe of turbulent motions. We found that the volume of WNM sampled here ($5.2\times10^5$\,pc$^3$), located at the outer edge of the Local Bubble, shows thermal properties in accordance with expected values for heating and cooling processes typical of the Solar neighbourhood: $P_{\rm th}/k_B = (4.4\pm2.6)\times 10^3$\,K\,cm$^{-3}$, $n=0.74\pm0.41$\,cm$^{-3}$, and $T_k=(6.0\pm1.3)\times 10^3$\,K. The WNM has the properties of sub/trans-sonic turbulence, with a turbulent Mach number at the largest scale probed here ($l=130$\,pc) of $\mathcal{M}_s = 0.87 \pm 0.15$, a density contrast of $\sigma_{\rho/\rho_0} = 0.6 \pm 0.2$, and velocity and density power spectra compatible with $k^{-11/3}$.  The low Mach number of the WNM provides dynamical conditions that allows the condensation mode of thermal instability (TI) to grow freely and form CNM structures, as predicted by theory. 
\end{abstract}

 

\section{Introduction} \label{sec:introduction}

Like for the Universe as a whole, the hydrogen atom is the most abundant element in galaxies, being in neutral (\HI), ionized (\HII) or molecular (H$_2$) form. The neutral phase is of primary interest for understanding star formation in galaxies; most of the mass and the volume of the interstellar medium is dominated by the diffuse phases, and the \HI\-to-H$_2$ transition is a key step that leads to the formation of cold, dense and self-gravitating structures, where stars form. We know that this transition occurs in the CNM ($T_k\sim 50$\,K, $n\sim 50$\,cm$^{-3}$) but the physical processes that lead to the formation of these structures is still a matter of debate. One key element in understanding the formation of CNM clouds in galaxies seems to be related to the turbulent cascade acting in the precursor environment, the WNM ($T_k \sim 8000$\,K, $n \sim 0.3$\,cm$^{-3}$). 

It is now well established from a theoretical perspective that the \HI\ can be described as a multi-phase turbulent flow \citep[e.g.,][]{hennebelle_structure_2007,saury_structure_2014}.
In this picture, the CNM occupies only a few percent of the volume. It is the result of a condensation process occurring in the volume-filling WNM that behaves more like a classical compressible, isothermal turbulent flow. Numerical studies have shown that the outcome of the condensation process depends  on the specific properties of turbulence in the WNM \citep[e.g.,][]{seifried_2011,saury_structure_2014,bellomi_2020}. In fact, many have suggested that cold and dense interstellar clouds form by moderately supersonic compression of warm atomic gas streams \citep[e.g.,][]{hennebelle_warm_2008}. These studies show that strongly supersonic turbulence leads to dynamical times that are too short to leave enough time for the cooling and for the condensation process to be efficient. On the other hand, too weak turbulence does not provide enough density fluctuations and mixing to enable the formation of thermally stable cold structures. The other essential parameter is the pressure of the WNM; it has to be high enough to bring the gas in the thermally unstable range \citep{wolfire_multiphase_1995,wolfire_neutral_2003}. 

Theory and numerical experiments have been of tremendous help to understand this complex non-linear process but this knowledge rests, for now, on a relatively small number of observations.
In fact, many of the fundamental physical properties of the WNM remains elusive for now, like its exact kinetic temperature, volume filling fraction, thermal pressure and turbulent Mach number. Most of these quantities are deduced either by theory or indirectly by comparison with other phases. There is still an unresolved debate about how the WNM is distributed in interstellar space. Is-it the extended skin of cold structures \citep{mckee_theory_1977} or a widespread inter-cloud medium~? The question of its relation with ionized phases of the ISM (warm and hot) is still very much open \citep{cox2005}. 

Early on, the ubiquity of a large and spectrally smooth 21\,cm component, seen in emission but not in absorption, was reported. \cite{dickey1979} mentioned that this warm, ``not strongly absorbing", gas corresponds to probably up to 75\% of the column density at high Galactic latitudes. The fact that the CNM is subdominant in the mass budget of the \HI\ was confirmed later on: the CNM accounts for about 10-30\% of the mass in the diffuse ISM at high Galactic latitudes \citep{dickey1979,heiles_millennium_2003,haud_gaussian_2007,kalberla_properties_2018,murray_21-sponge_2018,murray_2020} and possibly more in the vicinity of molecular clouds \citep[40\% according to][]{nguyen2019}.

On the other hand \cite{dickey1979} could not say how much of the WNM is warm ($T_k>>1000$\,K) and lukewarm (a few hundred K), a topic still debated. They had realized that very long integration times would be required to detect this gas in absorption and, to this day, the detection of the WNM in absorption remains a great challenge. So far, the highest temperatures estimated from absorption measurements are in the thermally unstable range \citep{mebold1982,carilli_1998,kanekar_temperature_2003,dwarakanath2002,begum_thermally_2010,murray_21-sponge_2018, nguyen2019} even with an optical depth sensitivity of $\sigma_{\tau}\sim10^{-3}$ per channel. The highest temperature ($7200^{+1800}_{-1200}$\,K) was obtained by \cite{murray_excitation_2014} who could detect a WNM signature by stacking 19 absorption spectra.

In these conditions, up to now the estimate of the WNM properties has been done using 21\,cm emission profiles. Two techniques were used, one is based on absorption  measurements against radio sources where the comparison between the absorption and emission profiles in the vicinity of the source provides a reliable description of the narrow CNM components. The contribution from the cold gas can then be removed from the emission profile, providing an estimate of the WNM column density and line width \citep[e.g.,][]{heiles_millennium_2003}. The second method is to decompose emission spectra only, using a sum of Gaussians. That was done early on \citep[e.g.,][]{mebold_intercloud_1972} and more recently using fully sampled data cubes \citep{haud_gaussian_2007,kalberla_properties_2018}. The average Doppler line width of the WNM at high Galactic latitude found by these studies is $7.6-9.5$\,km\,s$^{-1}$. This method has the advantage of not being limited to lines of sights crossing radio sources and therefore can provide larger statistics of the WNM gas. Its main drawback is the difficulties related to the Gaussian decomposition \citep[see][hereafter M19, for a discussion of the many pitfalls]{marchal_rohsa:_2019}.

For both methods, the information extracted from the emission profiles is limited to the column density, velocity centroid and Doppler line width. The difficulty to estimate the length on the line of sight over which the WNM emission is coming from, and the difficulty to separate thermal and turbulent contributions of the line, make the determination of the WNM density, filling factor and turbulent Mach number elusive for now.

Nevertheless these quantities are essential to understand the ISM evolution. In fact the process by which the diffuse gas condense, increasing its density by several order of magnitudes, implying gathering matter over very large volumes, is central in the general evolution of galaxies \citep{cox2005}. The efficiency with which the CNM forms and the timescale over which these structure exist, depend significantly on the dynamical properties of the inter-cloud medium, especially the pressure and turbulent Mach number of the WNM. The current study is an attempt to extract more information out of 21\,cm emission data cubes, and bring new observational constraints on the dynamical properties of the warm gas in the Solar neighborhood.

The paper is organized as follows. In Sect.~\ref{sec:north-ecliptic-pole}, we present the data used in this work and the Gaussian decomposition performed to model its multiphase structure. 
In Sect.~\ref{sec:HI-components}, we analyze the thermal phases at local velocities, their spatial distribution along the line of sight, and volume filling factors.
In Sect.~\ref{sec:disentangling-thermal-turbulent}, we describe the methodology used to disentangle thermal and turbulent contributions from the Doppler line width of the local WNM.
Gas properties (thermodynamic and turbulent) are analyzed in Sect.~\ref{sec:gas-properties}.
Section~\ref{sec:thermal-instability} examines the static and dynamic scales of TI.
A summary is provided in Sect.~\ref{sec:summary}.

\section{\HI\ spectral data and decomposition}
\label{sec:north-ecliptic-pole}

\subsection{Data}
\label{subsec:general-description}

\begin{figure}[!t]
  \centering
  \includegraphics[width=\linewidth]{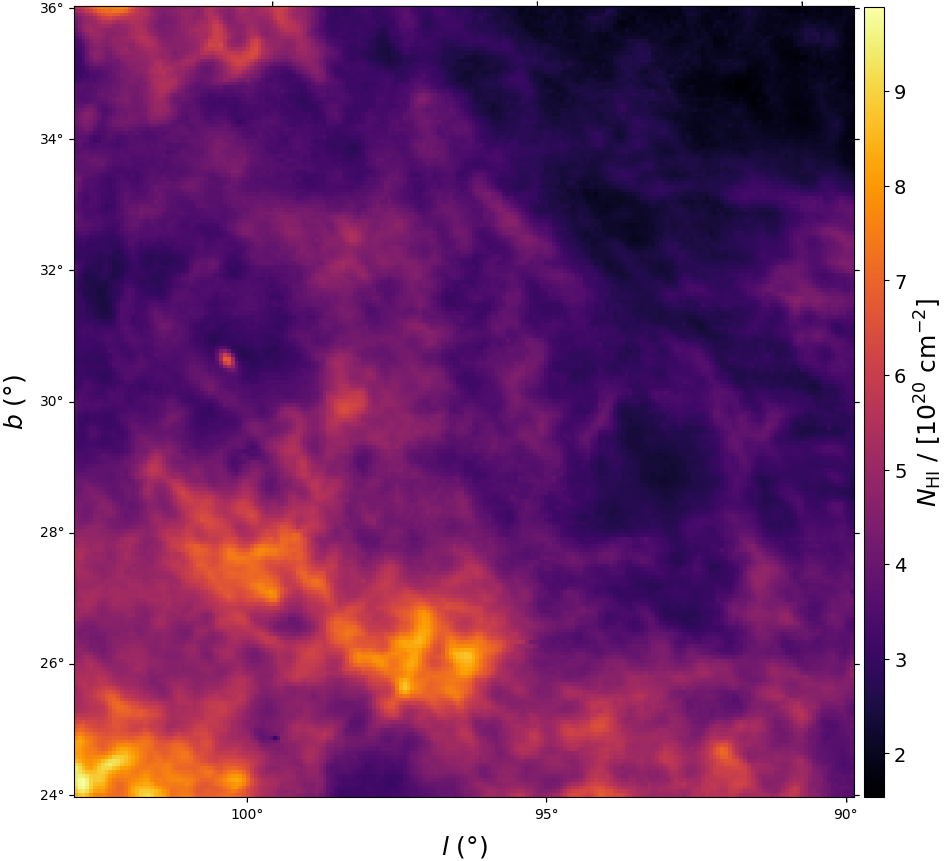}
  \caption{Column density map of the LVC and IVC gas observed in NEP.}
  \label{fig:NHI_TOT_NEP}
\end{figure}

\begin{deluxetable*}{lccccccccccccc}
\tablecaption{Mean kinematic properties (in \kms) of Gaussians inferred from NEP data}
\label{table::mean_var_NEP}
\tablewidth{0pt}
\tablehead{
\nocolhead{}  & \colhead{$G_1$} & \colhead{$G_2$} & \colhead{$G_3$} & \colhead{$G_4$} & \colhead{$G_5$} & \colhead{$G_6$} & \colhead{$G_7$} & \colhead{$G_8$} & \colhead{$G_9$} & \colhead{$G_{10}$} & \colhead{$G_{11}$} & \colhead{$G_{12}$} 
}
\startdata
$\langle\mub_n\rangle$ & -74.1 & -53.9 & -44.7 & -35.0 & -22.9 & -12.6 & -4.8 & -1.3 & 0.2 & 10.9 & 40.8 & 75.9 \\
$\langle\sigmab_n\rangle$ & 9.7 & 6.2 & 3.5 & 5.1 & 5.3 & 4.7 & 1.6 & 6.3 & 1.9 & 7.5 & 12.5 & 9.3 \\
\enddata
\end{deluxetable*}

The $12^\circ \times 12^\circ$ NEP field (or simply NEP) studied in this paper, located at $(l,\,b)=(96\fdg 40,\,30\fdg 03)$ was part of the GHIGLS\footnote{\url{http://www.cita.utoronto.ca/GHIGLS/}} \HI\ survey \citep{martin_ghigls:_2015} obtained with the Auto-Correlation Spectrometer (ACS) at the Green Bank Telescope (GBT).
NEP data have a channel spacing $\Delta v=0.807$\,\kms, an effective velocity resolution of about 1\,\kms and a spectral coverage $-365 < v\,[\kms] < 320$. 
The spatial resolution of the ACS data was about 9\farcm 4.
NEP was chosen from among other fields of the GHIGLS \HI\ survey to avoid the effect of velocity crowding and self-absorption, important at low latitudes. This is also the field with the largest spatial coverage, allowing us to probe the statistical properties of turbulence in the WNM over almost two order of magnitude in spatial scales.
Although there is a high velocity component (HVC) in NEP, it could be easily isolated in velocity. Only the intermediate velocity component (IVC) and the local velocity component (LVC) in the spectral range $-90 \leq v\,[\kms]\leq 90$ were kept for the purpose of this work. Figure~\ref{fig:NHI_TOT_NEP} shows the total column density map of NEP in this range.

\subsection{Gaussian decomposition}
\label{subsec:gaussian-decomposition}
We performed a multiphase separation of NEP using the publicly available code {\tt ROHSA}\footnote{\url{https://github.com/antoinemarchal/ROHSA}} \citepalias{marchal_rohsa:_2019}. {\tt ROHSA} is a multi-Gaussian fitting code whose function is to decompose hyper-spectral observations into a sum of spatially coherent components. In addition, {\tt ROHSA} was developed so that each resulting component has a similar velocity dispersion across the 2D position space, facilitating the association of each component to a given phase of the neutral ISM (WNM, LNM\footnote{LNM stands for Lukewarm Neutral Medium}, or CNM).

The decomposition used here is the one presented in Sect.~4. of \citetalias{marchal_rohsa:_2019}. The model $\tilde T_b\big(v_z, \thetab(\vb{r})\big)$ used to fit the measured brightness temperature $T_b(v_z, \vb{r})$ at a projected velocity $v_z$ and coordinates $\vb{r}$ is 
\begin{equation}
  \tilde T_b\big(v_z, \thetab(\vb{r})\big) = \sum_{n=1}^{N} G\big(v_z, \thetab_n(\vb{r})\big)
  \label{eq::model_gauss}
,\end{equation}
with $\thetab(\vb{r}) = \big(\thetab_1(\vb{r}), \dots, \thetab_n(\vb{r})\big)$ and where
\begin{equation}
  G\big(v_z, \thetab_n(\vb{r})\big) = \ab_n(\vb{r}) \exp
  \left( - \frac{\big(v_z - \mub_n(\vb{r})\big)^2}{2 \sigmab_n(\vb{r})^2} \right)
\end{equation}
is a Gaussian parametrized by $\thetab_n = \big(\ab_n, \mub_n, \sigmab_n\big)$ with
$\ab_{n} \geq \bm{0}$ being the amplitude, $\mub_{n}$ the position, and
$\sigmab_{n}$ the standard deviation 2D maps of the $n$-th Gaussian
profile across the plan of sky.
The parameters $\hat{\thetab}$ are obtained by minimizing the cost function described in \citetalias{marchal_rohsa:_2019}. It includes, for each parameter map $\big(\ab_n, \mub_n, \sigmab_n\big)$, a Laplacian filtering that penalizes the small spatial frequencies whose strength is controlled by a hyper-parameter. An additional term, minimizing the variance of $\sigmab_n$ is added to ensure the multiphase separation. {\tt ROHSA} was initialized with a sum of $N=12$ Gaussians to ensure a complete encoding of the signal with spatially coherent components and each hyper-parameters $\lambda_{\ab}$, $\lambda_{\mub}$, $\lambda_{\sigmab}$, and $\lambda'_{\sigmab}$ have been set to 1000. We refer the reader to \citetalias{marchal_rohsa:_2019} for a detailed discussion about the choice of the five (including $N$) user-parameters of {\tt ROHSA}. 

Mean velocity $\langle\mub_n\rangle$ and mean velocity dispersion $\langle\sigmab_n\rangle$ of the 12 Gaussian components $G_n$ are tabulated in Table~\ref{table::mean_var_NEP}. We refer the reader to \citetalias{marchal_rohsa:_2019} for a visualization of their column density maps, velocity fields, and dispersion velocity fields.

\section{HI components in the NEP field}
\label{sec:HI-components}

\begin{figure}[!t]
  \centering
  \includegraphics[width=\linewidth]{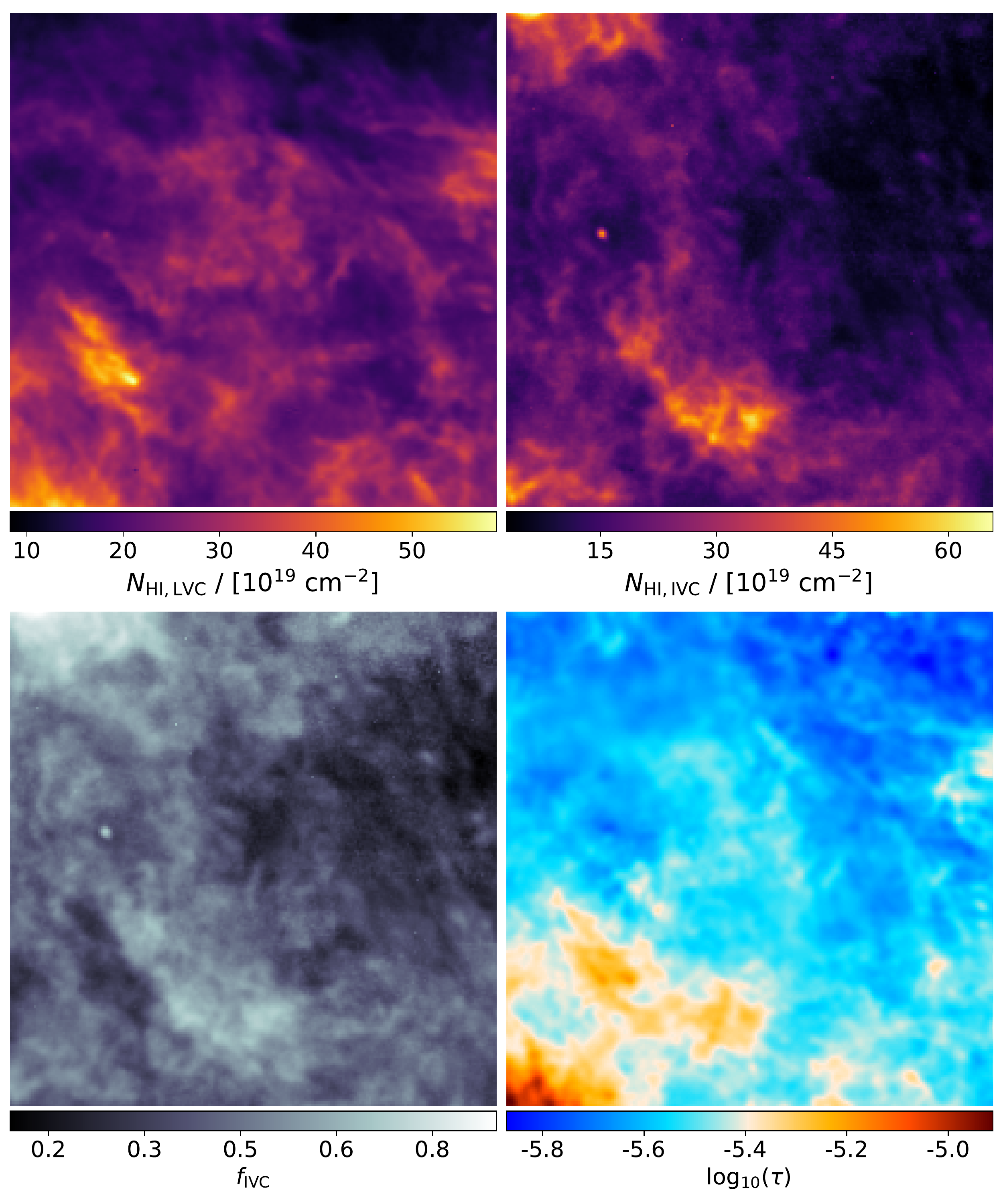}
  \caption{Top left: Column density map of the local gas in NEP. Top right: Column density map of IVCs. Bottom left: IVC mass fraction $f_{\rm IVC}$ map. Bottom right: Dust optical depth map from \cite{planck_collaboration_planck_2014}.}
  \label{fig:NHI_TOT_IVC_PLANCK_NEP}
\end{figure}

\begin{figure*}[!t]
  \centering
  \includegraphics[width=\linewidth]{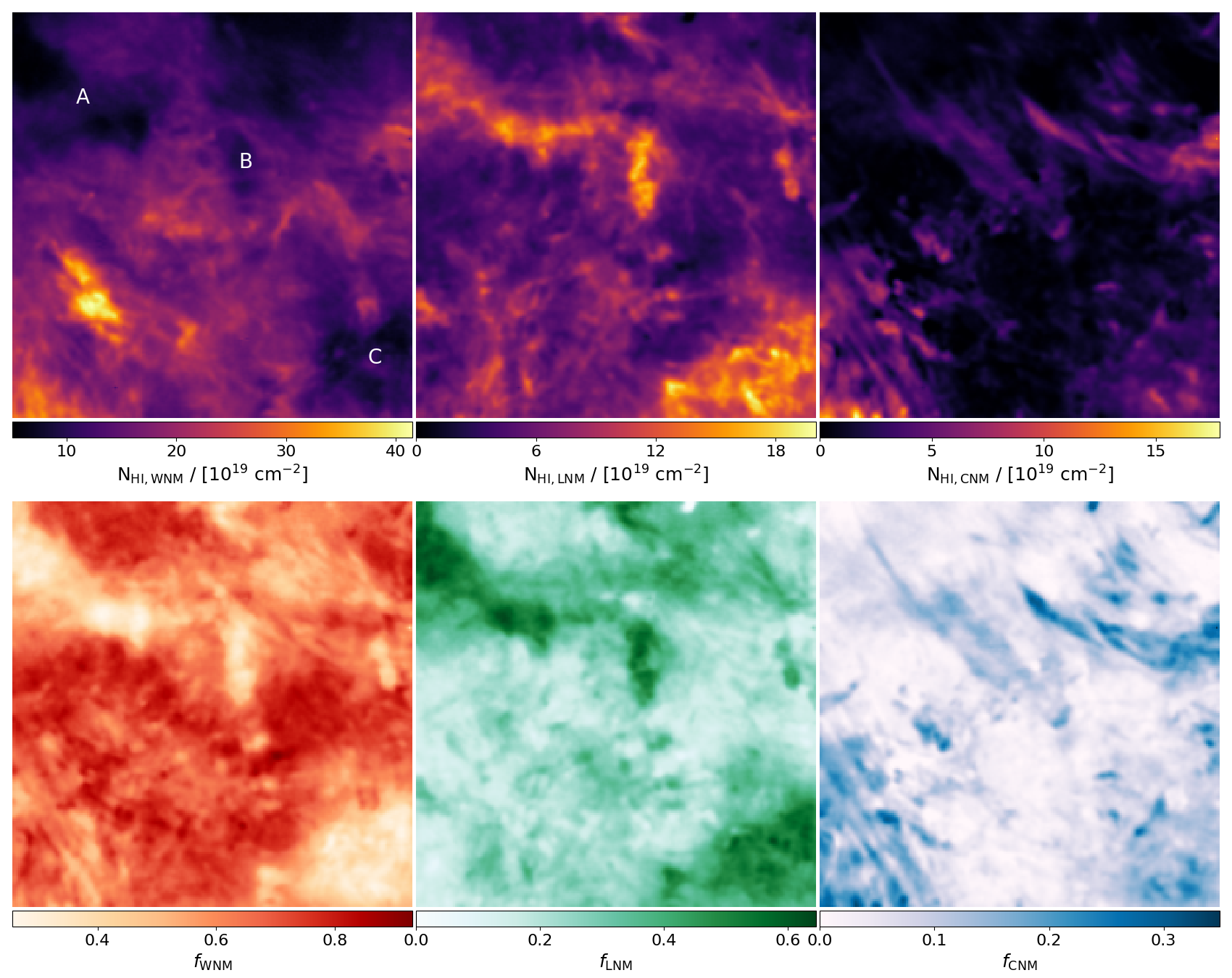}
   \caption{Top: Integrated column density $\nh^{\rm WNM}$, $\nh^{\rm LNM}$ and $\nh^{\rm CNM}$ fields of the three phase model. Bottom: $f_{\rm WNM}$, $f_{\rm LNM}$ and $f_{\rm CNM}$ column density fraction maps. Letters A, B and C indicated three anti-correlated regions of WNM and LNM discussed in the text.}
  \label{fig:NHI_and_fraction}
\end{figure*}

\subsection{Local and Intermediate Velocity Components}
\label{subsec:integrated_properties}

The Gaussian components extracted from NEP (see Table~\ref{table::mean_var_NEP}) can be separated in two groups; the first one ($\lvert\langle\mub_n\rangle\rvert > 20$ \kms) is composed of IVCs, the second one ($\lvert\langle\mub_n\rangle\rvert < 20$ \kms) represents the LVC. Figure~\ref{fig:NHI_TOT_IVC_PLANCK_NEP} shows column density maps of both components (top row) and the ratio of IVC column density fraction over total (bottom left). Many structures appear at all scales with significant variations of the column density over the field, from $\sim 1$ to $6\times 10^{20}$\,cm$^{-2}$. Interestingly the average and variations of the column density are very similar for the LVC and IVC components. These variations are directly reflected in the $f_{\rm IVC}$ map.

The dust optical depth at 353\,GHz (i.e., 850\,$\mu$m), $\tau_{353}$ \citep{planck_collaboration_planck_2014}, also shown in Fig.~\ref{fig:NHI_TOT_IVC_PLANCK_NEP} (bottom right), has a structure very similar to the total (LVC + IVC) integrated \HI\ emission (see Fig.~\ref{fig:NHI_TOT_NEP}). The average column density deduced from the dust optical depth at 353\,GHz is about $\langle N_{\rm H}^{\tau 353}\rangle \simeq 6.0 \times 10^{20}$\,cm$^{-2}$, 
assuming $\tau_{353}/N_{\rm H} = 6.3\times 10^{-27}$ of \cite{planck_collaboration_planck_2014}. This value is identical to the value deduced from the total 21\,cm integrated emission (sum of the IVC and LVC): 
$\langle \nh^{\rm tot} \rangle=6.0 \times 10^{20}$\,cm$^{-2}$. 
This implies that the IVCs, which constitutes about 40\% of the \HI\ column density in NEP, contain dust with a similar dust-to-gas ratio than the local gas, like what was found by \cite{collaboration_planck_2011}.

\subsection{Thermal phases at local velocities}
\label{subsec:integrated_properties}

\begin{deluxetable}{lcccccc}
\tablecaption{Mean properties of column density map for the local velocity component.}
\label{table:NHI-NEP}
\tablewidth{0pt}
\tablehead{
\nocolhead{}  & \colhead{$\langle \nh \rangle$} & \colhead{$\sigma_{\nh}$} & \colhead{$\sigma_{\nh}/\langle \nh \rangle$} & \colhead{$\langle f \rangle$} & \colhead{$\sigma_f$} \\
\nocolhead{}  & 10$^{18}$ cm$^{-2}$ & 10$^{18}$ cm$^{-2}$ & &   
}
\startdata
WNM & 146.5 & 51.3  & 0.35 & 0.64 & 0.13  \\
LNM & 56.9  & 30.1  & 0.53 & 0.28 & 0.11  \\
CNM & 13.5  & 18.2  & 1.35 & 0.08 & 0.06 \\
\enddata
\end{deluxetable}

\begin{figure}[!t]
  \centering
  \includegraphics[width=\linewidth]{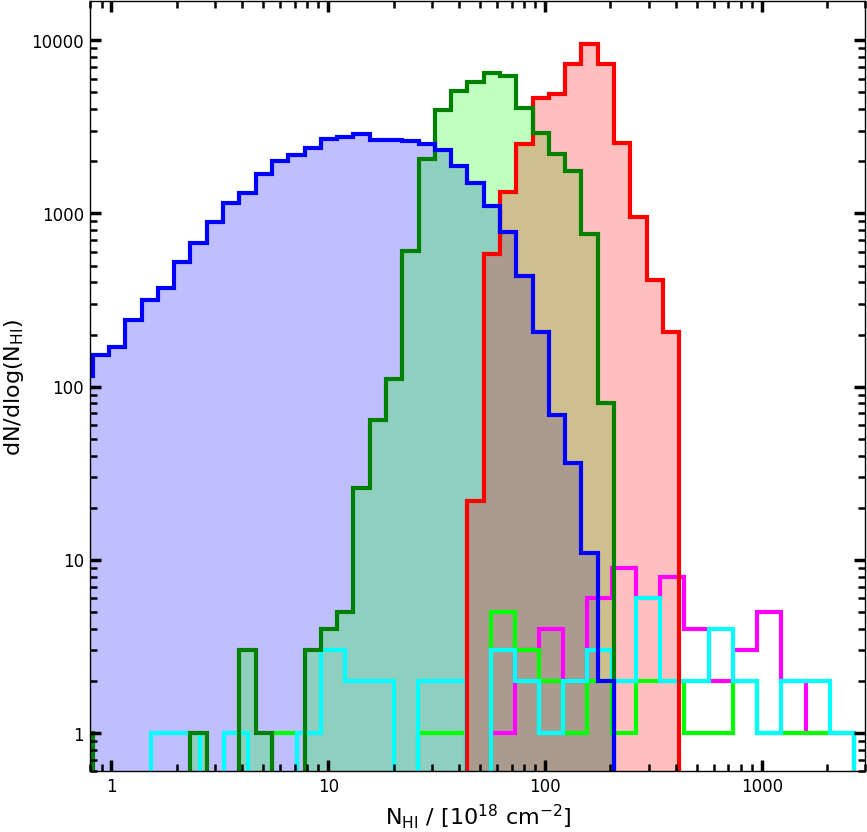}
 \caption{Probability distribution function of the column density of CNM (blue), LNM (green) and WNM (red). 
 Probability distribution function of WNM, LNM and CNM inferred from the 21-Sponge survey \citep{murray_21-sponge_2018} are shown in magenta, lime and cyan, respectively.
 }
  \label{fig:PDF_NHI}
\end{figure}

For the purpose of our analysis dedicated to the study of the WNM, we went a step further in the analysis of the result of the Gaussian decomposition and produced a multi-phase view of the LVC component. To do so we combined the Gaussian components at LVC velocities into three groups of similar width: WNM, LNM, and CNM. 
Specifically, Gaussian $G_{8}$ and $G_{10}$ are added to obtain the WNM, Gaussian $G_6$ forms the LNM and Gaussian $G_7$ and $G_9$ are combined to form the CNM (see Table~\ref{table::mean_var_NEP}).

Integrated column density fields of each phases are shown in Fig.~\ref{fig:NHI_and_fraction} (top row) as well as the mass fraction with respect to the total LVC column density (bottom row)
The column density maps of the WNM and LNM are anti-correlated in some areas, three in particulars that we have annotated A, B and C (see Fig.~\ref{fig:NHI_and_fraction}, top left). At each of these three positions, the low WNM column density areas seem to correspond to elongated structures in the LNM. These features appear clearly in the mass fraction maps. On the other hand, we note that the column density map of the CNM, composed of clumpy and filamentary structures, does not appear to be reflected in any other phases, except maybe for region C. Interestingly, in this region the CNM structures seem to have a smaller spatial extension than LNM structures. 

\begin{figure}[!t]
  \centering
  \includegraphics[width=\linewidth]{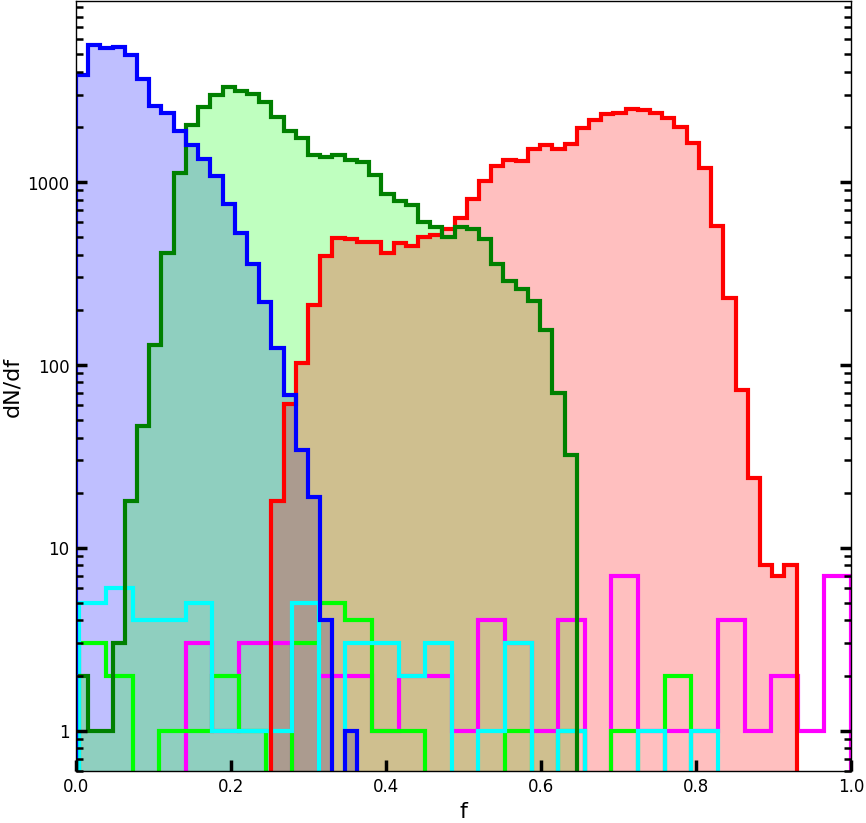}
 \caption{Like Fig.~\ref{fig:PDF_NHI} but for the mass fraction.}
  \label{fig:PDF_mass_fraction}
\end{figure}

The one-point probability distribution functions (PDFs) of the column density and mass fraction, for the three phases, are shown in Figs.~\ref{fig:PDF_NHI} and \ref{fig:PDF_mass_fraction}, respectively. Average and standard deviation values of each quantity are tabulated in Table~\ref{table:NHI-NEP}.
Each phase shows a rather complex column density distribution that is not particularly well described by a log-normal.
The median column density of the whole LVC is $2.25 \times 10^{20}$\,cm$^{-2}$. It is dominated by the WNM component that is 2.5 times larger than the the LNM and 10 times larger than the CNM.
On average, 64\% of the mass is in the WNM, 28\% in the LNM and only 8\% is in the CNM. However, as Table~\ref{table:NHI-NEP} shows, the standard deviation of these map is relatively high (35\%, 53\% and 135\% of the median value for the WNM, LNM and CNM, respectively). This reflects the important variations seen in Fig.~\ref{fig:NHI_and_fraction} (bottom row) and Fig.~\ref{fig:PDF_mass_fraction}, especially strong for the CNM that is very intermittent spatially, with a significant fraction 
of the field (55\%) with a mass fraction below 1\%. 

Such a spatial variability of the column density and mass fraction in each phase is to be expected 
because of the dynamical nature of TI. The values found in NEP are just one instance of the possible distributions.
Interestingly, the rather larger variability of the mass fractions observed here compares well with values that are found all over the sky with absorption surveys. As shown in Figs.~\ref{fig:PDF_NHI} and \ref{fig:PDF_mass_fraction}, for each phase, the ranges of $\nh$ and $f$ found in NEP are included in the broader distributions deduced from the 21-Sponge survey data by \cite{murray_21-sponge_2018}. 
On the other hand, we note that some $\nh$ values reported by \cite{murray_21-sponge_2018} are significantly larger than the maximum values found in NEP, but most of them are found at lower Galactic latitudes than NEP, where the lines of sight are longer. Therefore, because of the different line of sight lengths sampled in the two studies, the comparison of the column density statistics has a limited value. On the other hand the mass fraction of the different phases is something that should be less dependant on the length on the line of sight\footnote{$f_{\rm CNM}$ does not vary with latitude in a plane parallel model with two different scale heights for the CNM and the WNM.}. We found that the CNM mass fraction does not exceed 0.35 in NEP while it reaches more than 0.8 in 21-Sponge. We note that 61\% of the $f_{\rm CNM}$ 21-Sponge values larger than 0.35 are at lower Galactic latitudes than NEP but in this case the latitude can not explain this discrepancy. On the other hand, local effects like the Local Bubble where the CNM fraction is likely to be smaller than the Galactic average could explain the difference seen here.

The comparison with the results of \cite{murray_21-sponge_2018} highlights the fact that NEP has a low CNM fraction which makes it a particularly well suited field to study the properties of the diffuse inter-cloud medium. 

\subsection{Spatial distribution of the gas along the line of sight}
\label{sec:spatial_distribution}

\begin{figure}[!t]  
  \centering
  \includegraphics[width=\linewidth]{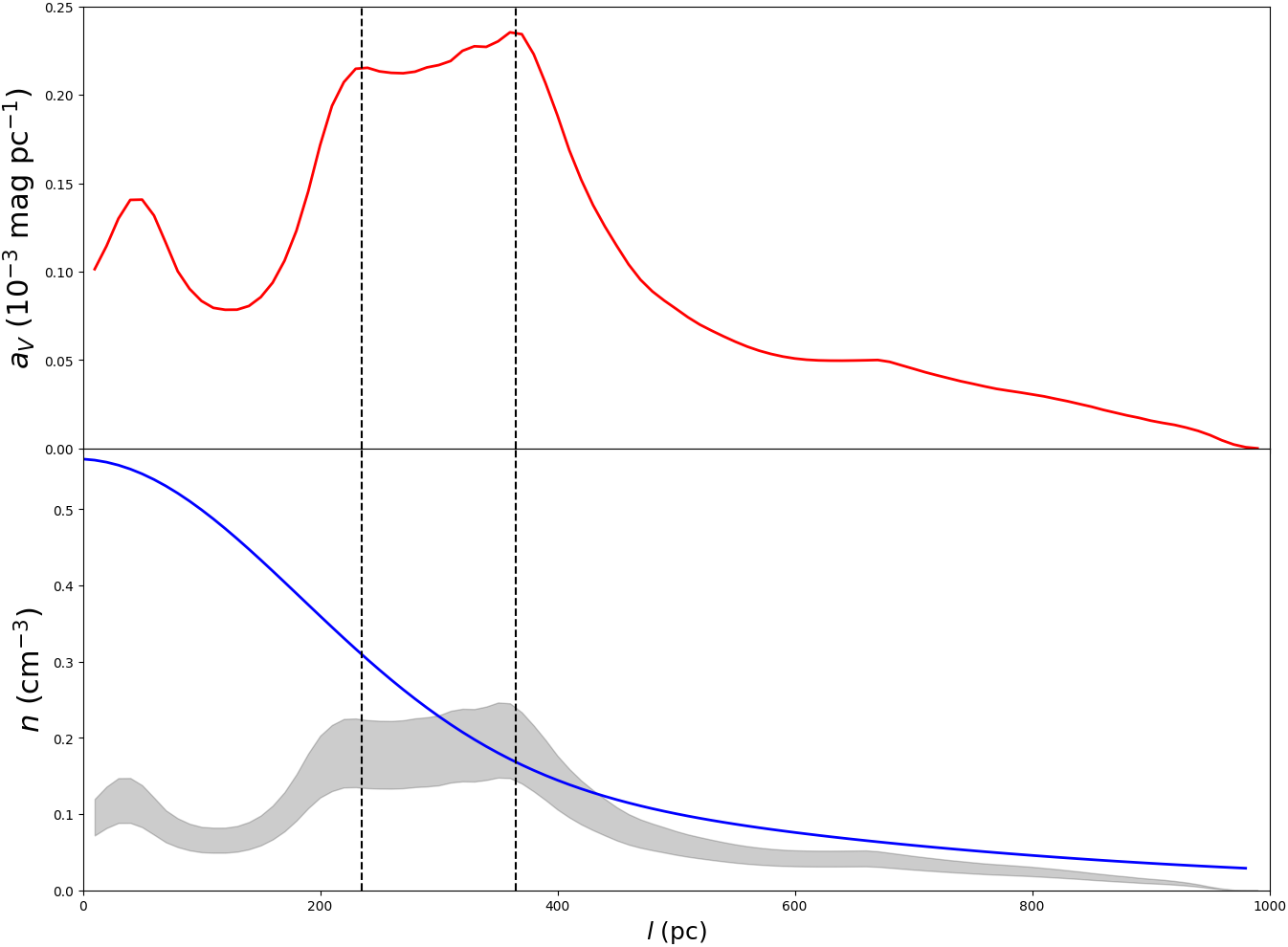}
  \caption{Top: Mean extinction profile $a_V$/pc in NEP based on the 3D extinction dust map performed by \cite{lallement_2019}. Bottom: Mean \HI\ density profile in NEP for an $N_{\rm H}$/E(B-V) ratio between 6-10$\times$10$^{21}$ (gray), and total \HI\ local density profile from \cite{dickey_h_1990} (blue). Vertical dashed lines indicate the size of the plateau, used as the typical depth $L_z$ of the fluid along the line of sight.}
  \label{fig:Av_LAL_NEP_mean}
\end{figure}

In order to make a quantitative study of the physical properties of the local \HI\ in NEP, one needs to estimate the distance and depth of the emitting gas. 
The decomposition of the 21\,cm data with {\tt ROHSA} enables a deblending of the LVC and IVC components, as well as a separation of the different thermal phases of the LVC, but for such high Galactic latitude fields, the velocity information provided by the 21\,cm data can not be used to estimate a kinematic distance.
In order to put constraints on the distance and depth of the gas, we relied on 3D tomography.

For the last two decades, and even more so now that the {\it Gaia} data are available, significant efforts have been put in mapping the interstellar medium in 3D using tomography techniques \citep{vergely_spatial_2010, lallement_local_1995, lallement_3d_2003, lallement_three-dimensional_2018, lallement_2019, green_measuring_2014, green_2018, rezaei_kh._inferring_2017}. Of these studies, the only one that mapped the local atomic ISM directly is \cite{vergely_nai_2001} who applied a regularized Bayesian inversion method on interstellar Ly$\alpha$ absorption measurements toward 454 stars. That way \cite{vergely_nai_2001} could produce a 60\,pc resolution map of the 3D density distribution of the neutral hydrogen within 250 pc of the Sun. Even with this coarse spatial resolution, their 3D map gives us an idea of how the \HI\ is distributed in the direction of NEP. A large cavity (the Local Bubble) is observed around the Sun with a volume density of about 0.1\,cm$^{-3}$. This cavity seems to extend over 200\,pc, the limit reached by this study.
 
No direct 3D mapping of \HI\ has been performed beyond 250\,pc. However, it is possible to approach it using a 3D dust extinction map. Using the same formalism as \cite{vergely_nai_2001}, \cite{lallement_2019} used {\it Gaia} DR2 photometric data combined with 2MASS to derive extinction measurements toward stars in a $6\times 6 \times 0.8$\,kpc$^3$ volume around the Sun. Here we use this product to compute the mean dust extinction per parsec profile, $a_V$, in the direction of NEP (see Fig.~\ref{fig:Av_LAL_NEP_mean}-top). The presence of the Local Bubble seen in \cite{vergely_nai_2001}, appears clearly in this profile. From 150 to 250\,pc, there is a smooth increase of the dust extinction per parsec, also seen in the $\nh$ distribution of \cite{vergely_nai_2001} (see their Fig.~6, top left). This makes it difficult to estimate a clear frontier for the Local Bubble. Beyond this first wall, $a_V$ remains rather constant for $\sim$130\,pc and then starts to decrease smoothly, on a one hundred parsec scale. 

From the $a_V$ profile, one can infer a gas volume density using the $N_{\rm H}/E(B-V)$ ratio 
\begin{equation}
    n_{\rm H}(l) = \frac{a_V(l)}{R_{\rm V}} \times \frac{N_{\rm H}}{E(B-V)}\,.
    \label{eq:density-from-Av}
\end{equation}

Recently, the classical value of $N_{\rm H}/E(B-V) = 5.8\times 10^{21}$\,cm$^{-2}$\,mag$^{-1}$ estimated by \cite{bohlin_survey_1978} has been the subject of significant revisions. Using different techniques, \cite{planck_collaboration_planck_2014,liszt_nh_2013,lenz_new_2017,zhu_2017,nguyen_2018} all obtained significantly higher values in the range $8\pm 1 \times$10$^{21}$ cm$^{-2}$\,mag$^{-1}$.

To visualize a realistic range of possibilities, we present in Fig.~\ref{fig:Av_LAL_NEP_mean}-bottom the $n_{\rm H}(l)$ density profile deduced from dust extinction, assuming $R_V=3.1$ and considering a $N_{\rm H}/E(B-V)$ ratio between 6-10$\times$10$^{21}$ cm$^{-2}$\,mag$^{-1}$. For the sake of comparison, we over-plot the \HI\ density profile from \cite{dickey_h_1990} that describes the average \HI\ volume density profile, $n(z)$, in the inner Galaxy, assuming cylindrical symmetry. This average \HI\ model and the model computed from the dust extinction curve appear to be roughly consistent in the range $l\sim$200-400\,pc where most of the local \HI\ mass is. This agreement is reassuring but one should not expect a perfect match as the \cite{dickey_h_1990} model does not take into account the particular conditions present in the solar neighbourhood, i.e. the Local Bubble.

At this point it is interesting to compare the column density computed by integrating the dust extinction curve 
\begin{equation}
    N_{\rm H}^{A_V} = \int n_{\rm H}(l) \, dl \,,
    \label{eq:NH_AV}
\end{equation}
to the column density inferred from the 21\,cm line emission. Integrating Eq.~\ref{eq:NH_AV} using the profile $a_V$ shown in Fig.~\ref{fig:Av_LAL_NEP_mean}, and assuming $N_{\rm H}/E(B-V)=8\times10^{21}$\,cm$^{-2}$\,mag$^{-1}$, we obtain $N_{\rm H}^{A_V}=2.4\times10^{20}$\,cm$^{-2}$. This is significantly lower than the average column density estimated from the dust optical depth or the total \HI\ emission~: $N_{\rm H}^{\tau353} = \nh^{\rm tot} = 6.0 \times 10^{20}$\,cm$^{-2}$.
Because they are related to dust, both $N_{\rm H}^{\tau 353}$ and $N_{\rm H}^{A_V}$ trace the total amount of hydrogen atoms along the line of sight (\HI\, \HII and H$_2$). The fundamental difference between these two estimates is that $N_{\rm H}^{\tau 353}$ traces dust emission to infinity while $N_{\rm H}^{A_V}$ stops at $l\sim 1000$\,pc, limited by the {\it Gaia} sample. This difference could be explained if another dusty cloud would be present beyond the detection limit of \cite{lallement_2019}.

Interestingly it appears that the median integrated column density of the local gas (LVC) inferred from the 21\,cm line is $\nh^{\rm LVC}=2.3\times10^{20}$\,cm$^{-2}$, a value in very good agreement with the one deduced from the dust profile $a_V$ (relative difference of 7.4~\%). Note that even if we consider the lower and upper value of $N_{\rm H}/E(B-V)$ ($6-10\times10^{21}$\,cm$^{-2}$\,mag$^{-1}$), the relative difference between $N_{\rm H}^{Av}$ and $\nh^{\rm LVC}$ remains within about $\pm$30~\%. This suggests that the IVC in NEP is located beyond 1 kpc from the Sun (i.e., in the Galactic halo, at $z > 575$\,pc). 

Building on this, in the following we will assume that the $a_V$ profile provides a description of the distribution of the LVC component along the los. Then it becomes possible to estimate the physical sizes $L_x$ (longitude), $L_y$ (latitude) and $L_z$ (depth) of the LVC in the region of NEP. Using the size of the plateau delimited by two vertical dashed lines in Fig.~\ref{fig:Av_LAL_NEP_mean}, we approximate the typical depth of the \HI\ along the line of sight $L_z\sim130$\,pc. At $l=300$\,pc, halfway down $L_z$, the $12^\circ \times 12^\circ$ region on the plane-of-sky translate into $L_x=L_y\sim63$\,pc.

\subsection{Volume filling factors}
\label{sec:volume-filling}

\begin{figure}[!t]
 \centering
 \includegraphics[width=\linewidth]{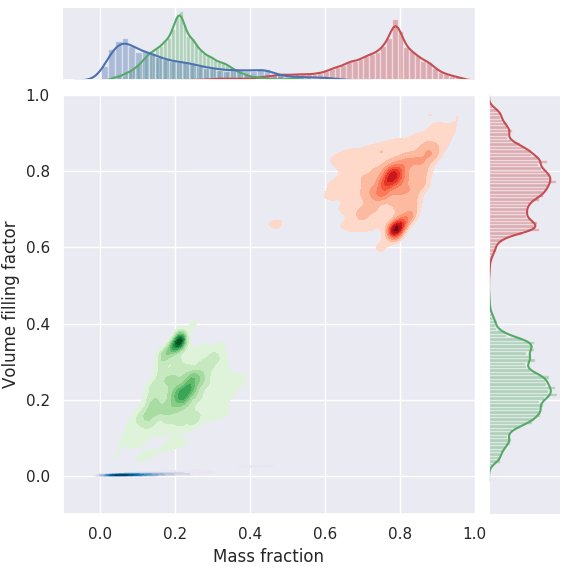}
 \caption{Two dimensional probability distribution function of the volume filling factor as function of the mass fraction for the WNM (red), LNM (green) and CNM (blue) along each line of light of a $10\times10\times40$\,pc$^3$ region with moderate CNM fraction (same region as in \citetalias{marchal_rohsa:_2019} of the numerical simulation of thermally bi-stable turbulence performed by \cite{saury_structure_2014}.)}
 \label{fig:filling_hist}
\end{figure}

Based on the previous 3D tomography analysis, we will now assume that the local velocity \HI\ component is coming from a slab of 130\,pc deep along the line of sight. The question now is how are the thermal phases distributed in this volume $\mathcal{V}$, in other words what is the volume filling factor of each phase. 

To explore this it is useful to turn to numerical simulations, like the one of \cite{saury_structure_2014}.  Figure~\ref{fig:filling_hist} shows the two dimensional PDF of the mass fraction and the volume filling factor for the WNM (red), LNM (green) and CNM (blue) (see top and right distributions for their respective one-point PDFs) for a $10\times10\times40$\,pc$^3$ cube, the same one used in \citetalias{marchal_rohsa:_2019} (see their Sect.~3). 
This specific simulation represents one realization of a thermally bi-stable \HI\ fluid. The mass fraction in each phase depends on the initial conditions and on the properties of the turbulent forcing. It appears that this one reproduces rather well the mass fractions observed in NEP (see Figs.~\ref{fig:PDF_mass_fraction} and \ref{fig:filling_hist}-top).

Because of the fact that the CNM has a density about 100 times larger than the WNM, it occupies a very small part of the volume. In the case of NEP where the CNM mass fraction is low, $f_M^{\rm CNM}= 0.08$, the CNM volume filling factor is likely to be $f_{\mathcal{V}}^{\rm CNM} \sim 0.001$. 
Therefore, most of the volume is filled by the low density LNM and WNM thermal components, a property shared with the numerical simulation used here. 

From the numerical simulation, we note that the LNM and WNM mass fractions and volume filling factors are correlated (see Fig.~\ref{fig:filling_hist}); on average, a larger mass fraction implies a larger volume filling factor. We also note that the structure of the LNM and WNM in the $f_M-f_{\mathcal{V}}$ plane are anti-correlated. This reflects the fact that these two thermal phases are dynamically linked and that their average density are similar. 
This link between the LNM and WNM translates into an anti-correlation in the column density as well \citep{saury_structure_2014}, something that is seen in NEP (see regions A, B and C in Fig.~\ref{fig:NHI_and_fraction}). This opens the possibility to use the mass fraction of the LNM and WNM as an indicator for their respective volume filling factor. From this, and neglecting the volume occupied by the CNM, we conclude that the LNM and WNM fill about 30\% and 70\% of the volume of the \HI\, respectively (see Table~\ref{table:NHI-NEP}). 

This coarse estimate of $f_{\mathcal{V}}$ was obtained in the idealized context where the totality of the gas is neutral. In fact this is an upper limits as some fraction of the volume in the direction of NEP is likely to be occupied by the diffuse Warm Ionized Medium, WIM. 
The joint analysis of pulsar DMs and diffuse H$\alpha$ emission by \cite{gaensler_vertical_2008} \citep[see also][]{reynolds1977,berkhuijsen_filling_2006} revealed that the WIM is best described by a collection of ionized structures (as opposed to a continuous medium) filling about 10\% of the volume (increasing with Galactic altitude) and with a global scale height of 1830$_{-250}^{+120}$\,pc. 
In this context it is difficult to evaluate precisely the fraction of the volume occupied by the WIM in the 130\,pc slab we are considering here. In addition we note that the correspondence between $N_{\rm H}^{\tau 353}$ and $\nh^{\rm tot}$ suggests that the column density of ionized gas $N_{\rm HII}$ is low in the direction of NEP, or that the WIM is deficient in dust (distinction between these two possibilities remains impossible to date). Anyhow one should keep in mind that 10-15\% of the volume considered here could be filled with ionized gas.

\section{The Warm Neutral Medium - Disentangling thermal and turbulent observable properties}
\label{sec:warm-neutral}
The decomposition of the 21\,cm data and the grouping of the widest components at local velocities has allowed us to build a model of the data cube of the WNM emission originating from the Galactic disk in the direction of NEP. This contains information not only on the column density of the WNM but also on the gas temperature and on its velocity component along the line of sight. In what follows, building on the previous constraints on the depth of local \HI\, we expose how we analyse the modeled cube to extract information about the thermal and turbulent observable properties of the warm \HI\ phase.

\subsection{Observable}
\label{subsec:observable}
In emission, the observed 21\,cm brightness temperature at position $\rb$ and velocity $v_z$, $T_b(v_z,\rb)$, depends on the variations along the line of sight of the gas volume density, $\rho$, the component of the velocity field along the los, $v_z$, and on the kinetic temperature $T_k$ \citep{miville-deschenes_physical_2007}. In the optically thin approximation the dependence of $T_b(v_z,\rb)$ on these physical quantities can be expressed as:
\begin{equation}
    T_b(v_z,\vb{r}) \propto \int_0^{L_z} \rho(\vb{r},z) \, \exp(-\frac{(v_z'-v_z(\vb{r},z))^2}{2 \sigma_{\rm th}^2(\vb{r},z)}) \, \dd z\,,
    \label{eq:TB_definition}
\end{equation}
where the integral is over the depth $L_z$ of the medium.
Note that the thermal velocity dispersion $\sigma_{\rm th}$ also varies with position $z$ along the line of sight as it depends on the kinetic temperature field, $T_k(\vb{r},z)$. 

From this, the zeroth, first and second moments of the emission, the column density, $\nh(\vb{r})$, the velocity centroid, $C(\vb{r})$, and the velocity dispersion, $\sigma_{T_b}(\vb{r})$, can be expressed the following way:
\begin{equation}
    \nh(\vb{r}) = \int_0^{L_z} \rho (\vb{r},z) \, \dd z\,,
    \label{eq:NHI_definition}
\end{equation}
\begin{equation}
    C(\vb{r}) = \frac{1}{\nh(\vb{r})} \int_0^{L_z} \rho (\vb{r},z) \, v_{z}(\vb{r},z) \, \dd z\,,
    \label{eq:Centroid_definition}
\end{equation}
and
\begin{equation}
    \sigma_{T_b}^2(\vb{r}) = \frac{1}{\nh(\vb{r})} 
    \int_0^{L_z} \rho(\vb{r},z) \, v_{z}^2(\vb{r},z) \, \dd z \, - \, C^2(\vb{r})\, + \, \frac{k_B \overline{T_k}(\vb{r})}{m_{\rm H}}\,,
    \label{eq:Sigmav_definition}
\end{equation}
where $k_B$ is the Boltzmann constant, $m_{\rm H}$ is the hydrogen atom mass, and $\overline{T_k}(\vb{r})$ is the average kinetic temperature along the line of sight, weighted by density:
\begin{equation}
\overline{T_k}(\vb{r}) = \frac{1}{\nh(\vb{r})} \int_0^{L_z} \rho (\vb{r},z) \, T_k(\vb{r},z) \, \dd z\,.
\end{equation}

\subsection{Density contrast}
\label{subsec:density-contrast}

\begin{figure*}[!t]
   \centering
   \includegraphics[width=0.49\linewidth]{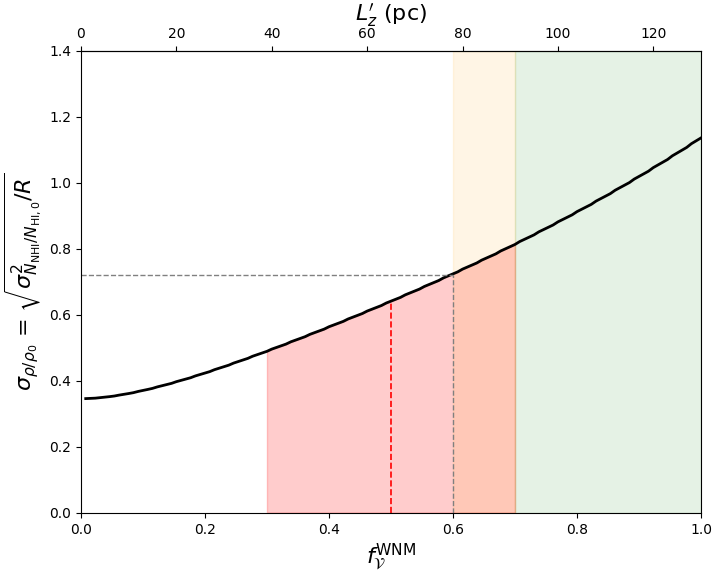}
   \includegraphics[width=0.49\linewidth]{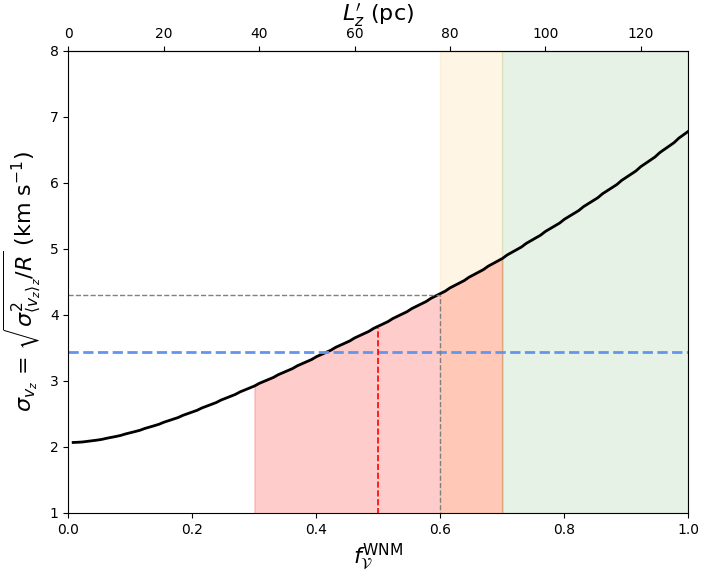}
   \caption{Left: Density contrast as a function of volume filling factor $f_{\mathcal{V}}^{\rm WMN}$ (or effective depth $L_z' = f_{\mathcal{V}}^{\rm WMN} \, L_z$ on the top axis). Green and orange areas show the volume occupied by the LNM and WIM, respectively. Right: Amplitude of turbulent motions as a function of $f_{\mathcal{V}}^{\rm WMN}$. Horizontal blue line shows the turbulent velocity dispersion obtained from the cloud-cloud velocity dispersion of CNM structures. The red dashed line and red area shows our final estimate of $f_{\mathcal{V}}^{\rm WMN}=0.5\pm0.2$.}
   \label{fig:Brunt}
 \end{figure*}

First, we analyse the density contrast of the WNM in NEP. To do so we used the formalism developed by  \cite{brunt_method_2010} to reconstruct the variance of a 3D physical field from 2D observation assuming only statistical isotropy of the underlying 3D physical field and knowing the physical scales (in 3D) of the fluid. 
This formalism is general enough to be applied to any kind of field, including density and velocity. First, we use this formalism to infer the density contrast $\sigma_{\rho/\rho_0}$ of the 3D density field from the column density contrast of its projection along the line of sight $\sigma_{N / N_0} = 0.346$. Using Parseval's Theorem, \cite{brunt_method_2010} showed that $\sigma_{N/N_0}$ and $\sigma_{\rho/\rho_0}$ are linked by the ratio 
\begin{align}
    R &= \left( \frac{\sigma_{N/N_0}}{\sigma_{\rho/\rho_0}} \right)^2 \\
    &= \frac{\left( \sum\limits_{k_x=-L_x/2+1}^{L_x/2} 
        \sum\limits_{k_y=-L_y/2+1}^{L_y/2}
        P^{3D}_{\rho}(k) \right) - P^{3D}_{\rho}(0)}{\left( \sum\limits_{k_x=-L_x/2+1}^{L_x/2} 
        \sum\limits_{k_y=-L_y/2+1}^{L_y/2}
        \sum\limits_{k_z=-L_z/2+1}^{L_z/2}
        P^{3D}_{\rho}(k) \right) - P^{3D}_{\rho}(0)}
    \label{eq:R}
\end{align}
where $P^{3D}_{\rho}(k)$ is the power spectrum of $\rho$. For a given field of size $L_x \times L_y$, two parameters control the ratio $R$: the slope of $P^{3D}_{\rho}(k)$, and the depth of the cube $L_z$ over which velocity fluctuations are averaged.

An important point has to be made here. Because of the fact that the WNM does not fill the volume fully, the scale over which the density fluctuations are averaged is not $L_z=130\,$pc but it is over an effective depth $L_z' = f_{\mathcal{V}}^{\rm WNM} \times L_z$. It is natural to expect, also due to $f_{\mathcal{V}}^{\rm WNM}$, that the power spectrum of the column density (projected 2D field), usually taken as a direct measure of the 3D density power spectrum, will in fact differ from $P^{3D}_{\rho}(k)\propto k^{-11/3}$ (see Sect.~\ref{subsec:inertial-range} and Appendix~\ref{app:volume-filling-Pk}). To account for this effect, we assumed $P^{3D}_{\rho}(k)\propto k^{-11/3}$, representative of a sub/trans-sonic turbulence. 

Assuming $P^{3D}_{\rho}(k)\propto k^{-11/3}$, $L_x=L_y=63$\,pc and $L_z=130$\,pc, the density contrast depends on the WNM filling factor. 
Figure~\ref{fig:Brunt} (left) shows the recovered value of $\sigma_{\rho/\rho_0}$ for the range 0 $< f_{\mathcal{V}}^{\rm WNM} <$ 1. 
Based on this analysis, the maximum possible value is $\sigma_{\rho/\rho_0} = 1.14$, but given our previous estimate of the WNM volume filling factor, $f_{\mathcal{V}}^{\rm WNM} \lesssim 0.6$, the contrast is likely to be $\sigma_{\rho/\rho_0} \lesssim 0.72$. Note that this estimate will be refined using further constrains (see Sect.~\ref{subsec:tracer-particles}) and $\sigma_{\rho/\rho_0}=0.6\pm0.2$ will be adopted in the following. 

\begin{deluxetable*}{lcccc}
\tablecaption{Mean properties of the local emission of the WNM in NEP\label{table:turbulence-NEP}}
\tablewidth{0pt}
\tablehead{
\nocolhead{\textbf{Quantity}} & \colhead{\textbf{Symbol}} & \colhead{\textbf{Value}} & \colhead{\textbf{Units}} }
\startdata
\hline
\textit{\textbf{Observable}} &&&                 
\\
Column density & $\langle\nh\rangle$ & 1.47$\times$10$^{20}$ & cm$^{-2}$ \\
Largest scale observed & $L_z$  & 130$\pm$20 & pc \\
Doppler velocity dispersion at $L_z$ & $\sigma_{T_b}$ & 8.0$\pm$0.6 & \kms \\
Volume filling factor & $f_{\mathcal{V}}^{\rm WNM}$ & 0.5$\pm$0.2 & \\
\hline
\textit{\textbf{Thermodynamic}} &&& \\
Thermal velocity dispersion & $\sigma_{\rm th}$ & 7.0$\pm$0.7 & \kms \\
Kinetic temperature & $T_k$ & (6.0$\pm$1.3)$\times$10$^3$ & K \\
Volume density & $n$ & 0.74$\pm$0.41 & cm$^{-3}$ \\
Thermal pressure & $P_{\rm th}$/$k_B$ & (4.4$\pm$2.6$)\times$10$^3$ & K\,cm$^{-3}$ \\
\hline
\textit{\textbf{Turbulent cascade}} &&& \\
Density contrast at $L_z$  & $\sigma_{\rho/\rho_0}(L_z)$ & 0.6$\pm$0.2 & \\
Turbulent velocity dispersion at $L_z$ & $\sigma_{v_z}(L_z)$ & 3.8$\pm$0.4 & \kms \\
Turbulent sonic Mach number at $L_z$ & $\mathcal{M}_s(L_z)$ & 0.87$\pm$0.15 & \\
Sound speed & $C_s$ & 7.67$\pm$0.82 & \kms \\
Turbulent velocity dispersion at 1\,pc & $\sigma_{v_z}(1)$ & 0.76$\pm$0.1 & \kms \\
Turbulent sonic Mach number at 1\,pc & $\mathcal{M}_s(1)$ & 0.17$\pm$0.03 & \\
Mean free path & $\lambda$ & (4.4$\pm$2.4)$\times$10$^{-4}$ & pc \\
Kinematic molecular viscosity & $\nu$ & (5.5$\pm$3.1)$\times$10$^{20}$ & cm$^2$\,s$^{-1}$ \\
Knudsen number at $L_z$ & $K_n$ & (3.4$\pm$1.8)$\times$10$^{-6}$ & \\
Reynolds number at $L_z$ & $Re$ & (3.3$\pm$1.8)$\times$10$^{5}$  & \\
Sonic scale & $\lambda_s$ & (1.9$\pm$1.1)$\times$10$^2$ & pc \\
Relative strength of solenoidal to compressive modes & $\zeta$ & 0.5$\pm$0.4 & \\
Dissipation scale & $\eta$ & (1.0$\pm$0.4)$\times$10$^{-2}$ & pc \\
Dissipation time & $t_{\eta}$ & 50$\pm$18 & kyr \\
Energy transfer rate & $\epsilon$ & (1.2$\pm$0.4)$\times$10$^{-4}$ & L$_{\odot}$\,M$_{\odot}^{-1}$ \\
Ambipolar diffusion scale & $l_{\rm AD}$ & (3.2$\pm$0.9)$\times$10$^{-2}$ & pc \\  
\hline
\textit{\textbf{Thermal instability}} &&& \\
Cooling time & $t_{\rm cool}$ & $\simeq$2.3 & Myr \\      
Dynamical time at sonic scale & $t_{\rm dyn}(\lambda_s)$ & 25$\pm$12 & Myr \\
Cooling length & $\lambda_{\rm cool}$ & $\simeq$17.5 & pc \\
Condensation criterion at sonic scale & $\mathcal{I}(\lambda_s)$ & 0.09$\pm$0.04 & \\
Field length & $\lambda_F$ & 0.10$\pm$0.06 & pc \\
\hline
\enddata
\end{deluxetable*}

\subsection{Thermal and turbulent velocity dispersions}
\label{sec:disentangling-thermal-turbulent}

The width of the emission line is given by Eq.~\ref{eq:Sigmav_definition}. In this equation, the first two terms on the right do not depend on $T_k$; they represent the second order moment of the $z$-component of the velocity field, weighted by density. The third term depends only on $T_k$; it represents the thermal broadening of the line. Eq.~\ref{eq:Sigmav_definition} is often written in a more compact form:
\begin{equation}
    \sigma_{T_b}^2 = \sigma_{v_z}^2 + \sigma_{\rm th}^2\,.
    \label{eq:sigmav}
\end{equation}
The Doppler line width $\sigma_{T_b}$ represents the width of the 21\,cm emission line. Equations~\ref{eq:TB_definition} to \ref{eq:Sigmav_definition} show that it is a quadratic sum of the thermal and turbulent motions of the gas along the line of sight, both weighted by density.
This observational mixture makes their separation impossible when considering the emission spectrum of a single line of sight. Therefore, a significant challenge in trying to extract the physical properties of turbulence of the WNM, for example its turbulent Mach number, is to estimate the thermal and turbulent contributions to the Doppler line width.
In order to do that, we investigated two independent methods. 
The first one uses the fluctuations of the centroid velocity field of the WNM $C(\vb{r})$ and
the second one uses the CNM components extracted with ROHSA as tracer particles of the WNM velocity field. 
For each method, we make the assumption that the statistical properties of the velocity field are isotropic (i.e., $v_{z}$ is statistically representative of the three directions). 

\subsubsection{Use of the centroid velocity field of the WNM}
\label{subsec:use-centroid-WNM}

\begin{figure}[!t]
  \centering
  \includegraphics[width=\linewidth]{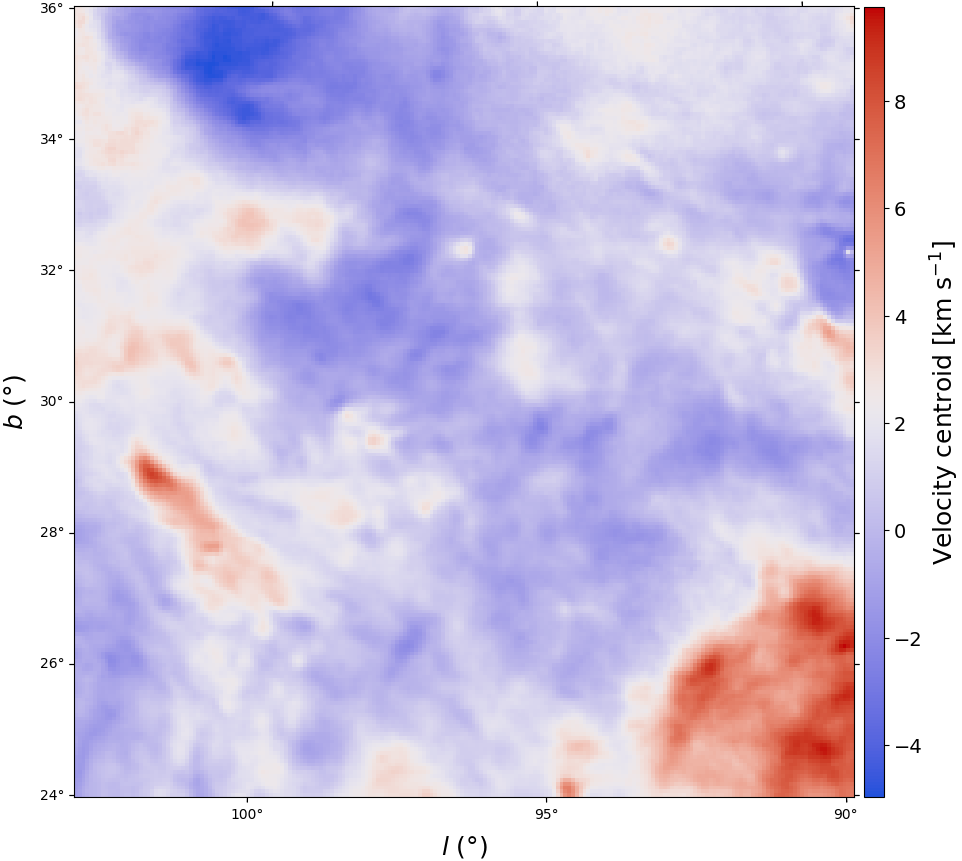}
  \caption{Centroid velocity field of the WNM at local velocities in NEP. 
  }
  \label{fig:CV_WNM}
\end{figure}

The $C(\vb{r})$ map of the WNM in NEP is shown in Fig.~\ref{fig:CV_WNM}. To our knowledge it is the first time that the centroid velocity of the WNM can be visualize. This map shows shallow fluctuations ($\sigma_{<v_z>_z} = 2.06$\,km\,s$^{-1}$).
As shown previously (see Eq.~\ref{eq:Centroid_definition}) the velocity centroid, $C(\vb{r})$, involves a complex combination of density and velocity fluctuations. Because of this mixture, using $C(\vb{r})$ to infer the statistical properties of $v_{z}(\vb{r},z)$ can only be done in very specific conditions. As shown by several studies \citep[e.g.,][]{miville-deschenes_use_2003,levrier_velocity_2004} the statistics of the 3D velocity field, $v_{z}(\vb{r},z)$, can be recovered from $C(\vb{r})$ when density fluctuations are small compared to the mean density of the fluid \citep[$\sigma_{\rho/\rho_0}\lesssim 0.5$, see][]{ossenkopf_interstellar_2006}. 
In Sect.~\ref{subsec:density-contrast} we found $\sigma_{\rho/\rho_0}\lesssim 0.7$. 
Therefore, we think that, in the very specific case of the WNM velocity field under study here, it is a fair approximation to use $C(\vb{r})$ as probe of $v_{z}(\vb{r},z)$. 

As stated in Sect.~\ref{subsec:density-contrast}, Eq.~\ref{eq:R} can also be applied on the centroid velocity field to infer the velocity dispersion of the 3D velocity field $\sigma_{v_z}$ from the velocity dispersion of its projection along the line of sight $\sigma_{\left<v_z\right>_z}$ = 2.06 \kms. Figure~\ref{fig:Brunt} (right) shows the recovered value of $\sigma_{v_z}$ for the range 0 $< f_{\mathcal{V}}^{\rm WNM} <$ 1. The maximum possible value is $\sigma_{v_z}= 6.8\,$km\,s$^{-1}$, but given our previous estimate of the WNM volume filling factor, $f_{\mathcal{V}}^{\rm WNM} \leq 0.6$, the turbulent contribution to the line width is likely to be $\sigma_{v_z} \leq 4.3\kms$.

\subsubsection{Use of CNM structures as tracer particles}
\label{subsec:tracer-particles}
To constrain further the amplitude of turbulent motions in the WNM we propose to use 
the cloud-cloud velocity dispersions $\sigma_{v_{z,c}}$ of CNM structures as being representative of the turbulent velocity field $\sigma_{v_z}$ of the inter-cloud medium. In this picture, the volume occupied by each dense CNM structure is very small compared to the volume in which they move. By analogy to fluid mechanics laboratory experiments, these CNM structures can be seen as tracer particles added to the fluid (WNM) to highlight the statistical properties of these turbulent motions.
This method makes the assumption that the CNM structures observed in NEP are homogeneously distributed in the $\mathcal{V}$.

To compute the velocity dispersion of CNM structures, we use the ensemble of Gaussians from components $G_7$ and $G_9$ (see Table.~\ref{table::mean_var_NEP}). We remind the reader that each Gaussian $i$ has three parameters, namely its amplitude $a_i$, its projected velocity along the line of sight $\mu_i$ and its velocity dispersion $\sigma_{v_i}$. The velocity dispersion of CNM structures $\sigma_{v_{z,c}}$ is defined as the square of the weighted variance of the velocity parameters of the Gaussian sample  
\begin{equation}
    \sigma_{v_{z,c}} = \sqrt{\frac{\sum_{i=0}^{N}{w_i} \left(\mu_i - \overline{\mu}^* \right)^2}{\sum_{i=0}^{N}{w_i}}}\,,
\end{equation}
where
\begin{equation}
    \overline{\mu}^* = \frac{\sum_{i=0}^{N}{w_i} \mu_i}{\sum_{i=0}^{N}{w_i}}\,,
\end{equation}
and,
\begin{equation}
    w_i = a_i \sigma_i\,.
\end{equation}
We find a velocity dispersion $\sigma_{v_{z,c}}$ = 3.5 \kms. Going back to the method presented in the previous section, if this value is representative of the WNM turbulent motion along the line of sight, it implies $f_{\mathcal{V}}^{\rm WNM}=0.43$ (see blue dotted line in Fig.~\ref{fig:Brunt} (right)).

Combining constraints from the centroid velocity field of the WNM and the cloud-cloud velocity dispersions of CNM structures we adopted values that are the average of the two methods, and uncertainties based on the limits set by them. Therefore we conclude that $\sigma_{v_z}$ = 3.8$\pm$0.4\,km\,s$^{-1}$, and $f_{\mathcal{V}}^{\rm WNM}$ = 0.5$\pm$0.2. Using this better estimate of $f_{\mathcal{V}}^{\rm WNM}$, we refine our estimate of the density contrast $\sigma_{\rho/\rho_0}=0.6\pm0.2$ (see Fig.~\ref{fig:Brunt} (left)). 

\section{Gas properties of the WNM in NEP} 
\label{sec:gas-properties}
A number of properties of the WNM in NEP – physical, thermodynamic, and turbulent as calculated in separate subsections below – are summarized in Table~\ref{table:turbulence-NEP}.

\subsection{Thermodynamic properties} 
\label{subsec:thermody}
\begin{figure}[!t]
  \centering
  \includegraphics[width=\linewidth]{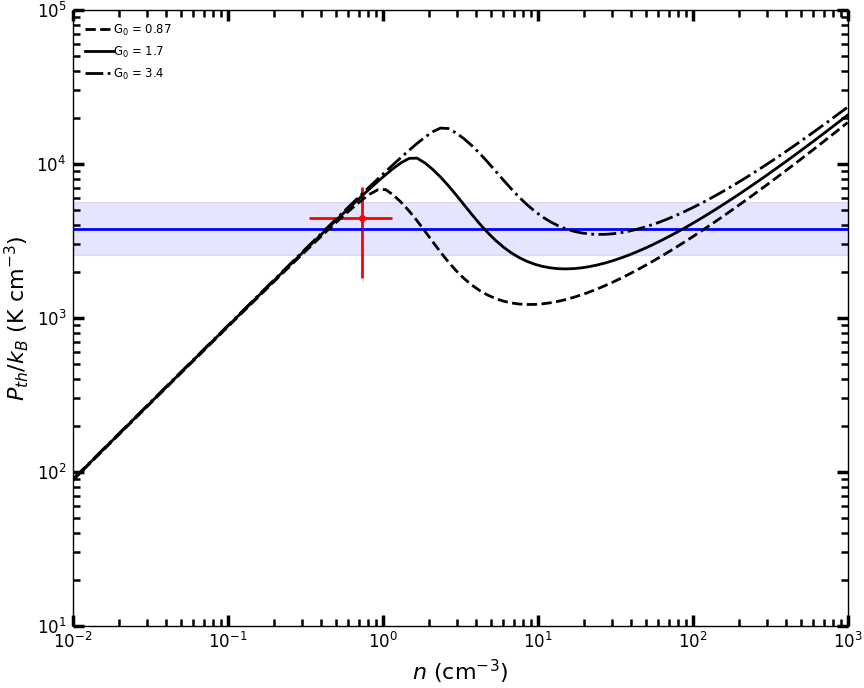}
  \caption{$P_{\rm th}/k_{B}$-$n$ diagram: Red cross shows the values obtained in the WNM of NEP. Standard model ($N_c$ = 1$\times$10$^{20}$\,cm$^{-2}$) of \cite{wolfire_neutral_2003} is over-plotted for three different values of the FUV interstellar radiation field strength $G_0=(0.84,1.7,3.4)$. Blue line shows the mean pressure of the cold neutral medium obtained by \cite{jenkins_distribution_2011}.}
  \label{fig:wolfire_NEP_cross}
\end{figure}

The kinetic temperature, volume density, and thermal pressure are
\begin{equation}
\label{eq:Tk}
    T_k = \frac{\sigma_{\rm th}^2 \, m_{\rm H}}{k_B}\,,
\end{equation}
\begin{equation}
    n = \frac{\langle \nh \rangle}{L_z'}\,,
\end{equation}
and
\begin{equation}
\label{eq:P}
    P_{\rm th}/k_B = n T_k\,,
\end{equation}
respectively. 
Subtracting $\sigma_{v_z}=3.8$\,km\,s$^{-1}$ from the observed mean velocity dispersion (see Eq.~\ref{eq:sigmav}), the mean thermal velocity dispersion is $\sigma_{\rm th}=7.0\pm0.8$ \kms. 
Using Eqs.~\ref{eq:Tk} to \ref{eq:P}, we find that $T_k=(6.0\pm1.3)\,\times 10^3$ K, $n=0.74\pm0.41$\,cm$^{-3}$, and $P_{\rm th}$/$k_B=(4.4\pm2.6)\times10^3$\,K\,cm$^{-3}$. 

\begin{figure*}[!t]
  \centering
  \includegraphics[width=0.49\linewidth]{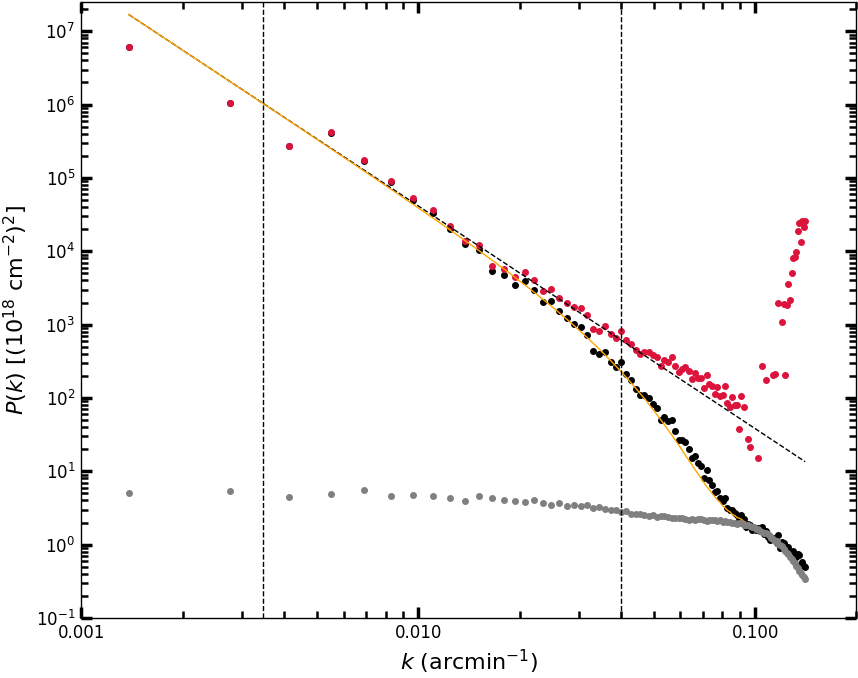}
  \includegraphics[width=0.49\linewidth]{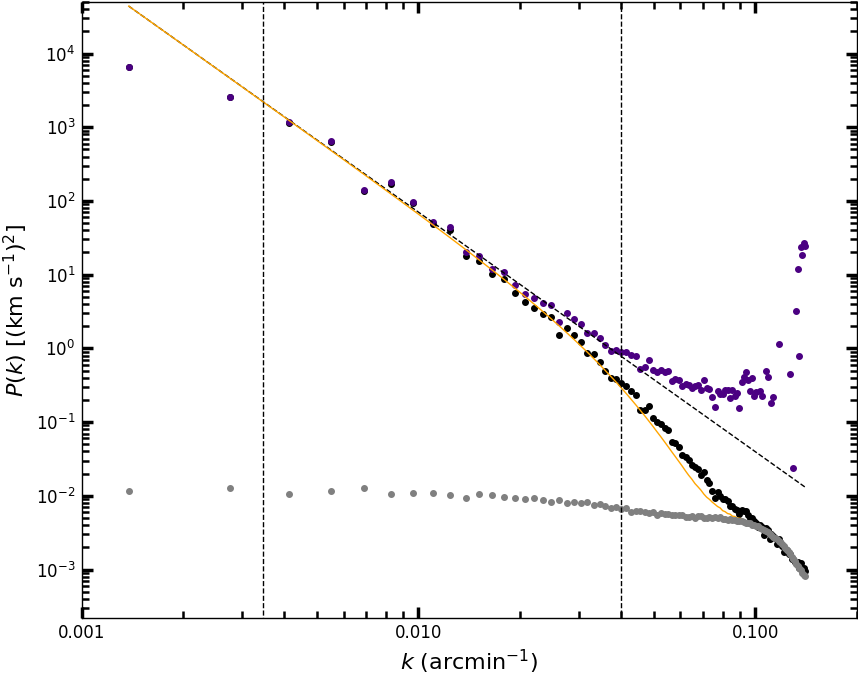}
  \caption{Spatial power spectrum model of the column density map (left) and centroid velocity field (right) of the WNM in NEP. $P(k)$ are given by the black dots. Grey points represent the noise component $N(k)$. Red points show the corrected power spectrum ($P(k) - N(k)$/$B(k)$). Dotted line shows the  model fits between the vertical dashed lines, chosen to select k ranges avoiding noise or systematics (high- and low-k limits, respectively). Orange line shows the result of the global modeling of Eq.~\ref{eq:model-Pk}. Exponents $\gamma_{N} = -3.04\pm0.08$ and $\gamma_{\left<v_z\right>_z} = -3.25\pm0.06$.}
  \label{fig:SPS_NHI_WNM_CV_WNM}
\end{figure*}

Figure~\ref{fig:wolfire_NEP_cross} shows a $P_{\rm th}/k_B-n$ diagram where the pressure and density inferred in this work is shown in red. 
For the sake of comparison, this figure also shows the range of pressure measured in the CNM at high Galactic latitudes by \cite{jenkins_distribution_2011} (blue line), as well as 
the standard equilibrium curves of \cite{wolfire_neutral_2003} computed for $N_c$ = 1$\times$10$^{20}$\,cm$^{-2}$ and three different values of FUV interstellar radiation field strength $G_0=(0.84,1.7,3.4)$ \citep[in units of the Draine (1978) field strength,][]{draine_1978}, representative of expected variations in the diffuse ISM.\footnote{These variations can be due to the distribution of stars in $\mathcal{V}$ which locally increase or decrease $G_0$, and/or variations due to the density field topology \citep{parravano_2003}. The latter induces a natural shielding of the radiation field, causing a decrease proportional to the column density of the gas.} 

We note that the WNM pressure estimated here is in complete agreement with the average value found in the local CNM by \cite{jenkins_distribution_2011} ($(3.8\pm1.5)\times10^3$\,K\,cm$^{-3}$). In addition, considering the realistic range of $G_0$ shown in Fig.~\ref{fig:wolfire_NEP_cross}, our results show that the values of pressure and density are compatible with a WNM in thermal equilibrium calculated with standard, solar neighbourhood properties for the cooling and heating mechanisms.
 
\subsection{Statistical properties of the turbulent energy cascade}
\label{sec:turbulence}

The separation of the thermal and turbulent contribution to the line width, as well as the estimate of the depth of the line of sight, allow us to quantify the statistical properties of the turbulent energy cascade in the WNM of NEP. 

\subsubsection{Inertial range}
\label{subsec:inertial-range}

The spatial power spectra of $\nh(\rb)$ and $C(\rb)$ shown in Fig.~\ref{fig:SPS_NHI_WNM_CV_WNM} (see Appendix~\ref{app:sps-procedure} for the procedure) are both well described by single power laws. This confirms the existence of an inertial range in NEP, over the projected spatial range of the observation from a few parsecs up to $L_x=L_y$.
Note that the depth along the line of sight, $L_z=130$\,pc, where the turbulent line width $\sigma_{v_z}(L_z)$ is measured, is significantly larger than the largest scale measured in projection on the sky $L_x=L_y=63$\,pc. From Fig.~\ref{fig:SPS_NHI_WNM_CV_WNM} alone it is impossible to ensure that an inertial range exist up to $l \sim L_z$. 
The back of the WNM layer studied here is at a distance of $\sim400$\,pc (see Fig.~\ref{fig:Av_LAL_NEP_mean}) which, at the latitude of NEP, corresponds to a height of about 200\,pc. The WNM volume under study here is thus within the WNM layer \citep[HWHM\,$\sim265$\,pc]{dickey_h_1990}. As it is likely that the outer scale of interstellar turbulence is of the order or larger than the disk thickness \citep{wolfire_neutral_2003}, in the following we will assume that the inertial range of turbulence goes up to $l \sim L_z$.

The slopes of the spatial power spectra of $\nh(\rb)$ and $C(\rb)$ are $-3.1\pm0.1$ and $-3.3\pm0.1$, respectively. 
\cite{martin_ghigls:_2015} performed a similar analysis using a multiphase decomposition of the emission of the local gas in NEP defined between $-20.5 < v_z\,(\kms) < 47.9$. 
They found an exponent $-2.7\pm0.1$ for the column density, slightly shallower than that obtained here.
Note however, that the WNM modeled by \cite{martin_ghigls:_2015} is a combination of our WNM and LNM phases which could explain this difference. 

The two slopes obtained here for $\nh(\rb)$ and $C(\rb)$ are flatter than the expected -11/3 value for compressible sub/trans-sonic turbulence \citep{kim_density_2005}. As noted in Sect.~\ref{subsec:density-contrast}, the effect of a partial volume filling factor of the WNM could explain that difference. To investigate this, we used fractional Brownian motions (fBms) simulations to which we applied partial volume filling factors (see Appendix~\ref{app:volume-filling-Pk}). We found that whatever the statistical properties of the masking field that simulate $f_{\mathcal{V}}^{\rm WNM} < 1$, the power spectrum slope of its projection is flatter than the true 3D value. We note though that the effect is less dramatic for the velocity centroid than for the column density field.

Therefore, from the power spectrum of the centroid velocity (or column density) itself, it is impossible to conclude on the nature of the turbulence, especially on its turbulent Mach number. Nevertheless we note that our results are compatible with $\gamma_{v_z} = \gamma_{N} = -11/3$, the expected value for compressible sub/trans-sonic turbulence. 

\subsubsection{Turbulent sonic Mach number and Reynolds number}

We quantify the strength of turbulence using the turbulent sonic Mach number and the turbulent Reynolds number. Note that these two are function of scale. The turbulent sonic Mach number is
\begin{equation}
    \mathcal{M}_s(L_z) = \frac{\sqrt{3} \, \sigma_{v_z}(L_z)}{C_s}
    \label{eq:mach}
\end{equation}
where the adiabatic sound speed is
\begin{equation}
    C_s = \sqrt{\frac{\gamma k_B T_k}{\mu m_H}},
    \label{eq:sound-speed-WNM-NEP}
\end{equation}
with $\mu$ = 1.4 is the molecular weight for the atomic Galactic composition, and $\gamma=5/3$ the adiabatic index of monoatomic gas. 
Using Eq.~\ref{eq:mach}, we find $\mathcal{M}_s(L_z)=0.87\pm0.15$, showing that the WNM in NEP is in a subsonic regime at scale $L_z$. This result is reminiscent of the low Mach number found for the WIM by \cite{gaensler_low-mach-number_2011}.

An important property of the turbulent velocity field is that it follows a scaling law dictated by the nature of the turbulent cascade and controlled by an exponent $q$. In a sub/trans-sonic regime, we expect $q=1/3$ (equivalent to -11/3 for the power spectrum scaling law). The so-called $\sigma_v-l$ relation is
\begin{equation}
    \sigma_{v_z}(l) = \sigma_{v_z}(1) \, l_{\rm pc}^{1/3}\,,
    \label{eq:sigma-v-NEP}
\end{equation}
where $ \sigma_{v_z}(1) = \sigma_{v_z}(L_z) L_z^{-1/3}$ is the velocity dispersion of the turbulent velocity field along the line of sight at 1\,pc, and $l_{\rm pc}$ is the scale in pc. 
Here we find that $\sigma_{v_z}(1) = 0.76\pm0.1$\,km\,s$^{-1}$. Interestingly this value is equivalent to the average found in much denser regions of the ISM; \cite{miville-deschenes_physical_2017} found $\sigma_{v_z}(1) \sim 0.8$\,km\,s$^{-1}$ for molecular clouds in the Solar neighbourhood. This echoes the insight of \cite{hennebelle_turbulent_2012} who mentioned that molecular clouds and the \HI\ are likely to be part of the same turbulent cascade. Similarly, combining Eqs.~\ref{eq:mach} and \ref{eq:sigma-v-NEP}, it follows 
\begin{equation}
    \mathcal{M}_s(l) = \mathcal{M}_s(1) \, l_{\rm pc}^{1/3}\,,
    \label{eq:mach_scale}
\end{equation}
that gives $\mathcal{M}_s(1)=0.17\pm0.03$.

In the following, we assume a mean free path $\lambda=1/\sigma n$ with $\sigma=1\times 10^{-15}$\,cm$^2$ for the hydrogen cross-section \citep{lequeux_milieu_2012} and a thermal velocity $v_{\rm th}=\sqrt{8/\pi}C_s$ to find the kinematic viscosity
\begin{equation}
    \nu = \frac{1}{3} \lambda v_{\rm th}
\end{equation}
Values of these properties are given in Table~\ref{table:turbulence-NEP}.

For an isothermal gas, the Knudsen number $Kn(L_z)=\lambda$/$L_z$ and $\mathcal{M}_s$ are linked directly to the Reynolds number
\begin{equation}
    Re(L_z) = \sqrt{\frac{\pi}{2}} \frac{\mathcal{M}_s(L_z)}{Kn(L_z)}.
    \label{eq:Reynolds-from-mach}
\end{equation}
which quantifies the ratio of advection and diffusion in a turbulent fluid.
We find $Re(L_z)=(3.2\pm1.4)\times10^{5}$ (see also Table~\ref{table:turbulence-NEP} for the corresponding Knudsen numbers), corresponding to a regime of fully developed turbulence at scale $L_z$ \citep{reynolds_experimental_1883}.

\subsubsection{The sonic scale}
\label{subsec:sonic-scale}
Using Eq.~\ref{eq:mach_scale}, the sonic scale is 
\begin{equation}
    \lambda_s = L_z \mathcal{M}_s^{-3}(L_z)\,,
    \label{eq:sonic-length}
\end{equation}
that satisfies $\mathcal{M}_s(\lambda_s)=1$. We find $\lambda_s=(1.7\pm1.1)\,\times10^2$\,pc. Above this scale, the turbulent cascade starts to be supersonic and is expected to be described by $q=0.5$ (equivalent $P(k) \propto k^{-4}$). 
A better sampling of high spatial scales could confirm a break in the scaling law of the projected velocity field and allow a parallel and complementary determination of $\lambda_s$. 

\subsubsection{Relative strength of solenoidal to compressive modes}

Following \cite{federrath_density_2008}, we estimate the relative strength of solenoidal to compressive modes $\zeta$ assuming 
\begin{equation}
    \sigma^2_{\rho/\rho_0} = b^2\mathcal{M}_s^2(L_z)
    \label{eq:mach-density}
\end{equation}
where $b = 1 - (2/3)\zeta$ for a 3D geometry. We find $\zeta=0.47\pm0.25$. 
This value is compatible with a mixture of solenoidal and compressive modes, close to the natural state for which the energy is naturally distributed between the two ($\zeta=0.5$). 

A method to estimate $\zeta$ from hyper-spectral observations was proposed by \cite{brunt_observational_2014} and recently applied in the Orion B molecular cloud by \cite{orkisz_turbulence_2017}. A similar application on NEP is beyond the scope of the paper but would provide a direct comparison to the value obtained here. 

\subsubsection{Energy transfer and dissipation}
Here we consider two mechanisms of energy transfer and dissipation along the turbulent cascade: the kinematic molecular diffusion, and the ambipolar diffusion.

The dissipation scale $\eta$, on which the smallest eddies dissipate the turbulent energy into heat through viscosity, is
\begin{equation}
    \eta = L_z Re(L_z)^{-3/4}\,.
    \label{eq:disspation-scale-NEP}
\end{equation}
We find $\eta=(9.8\pm3.3)\,\times10^{-3}$\,pc which is an order of magnitude higher than the mean free path of hydrogen atoms $\lambda$. The dissipation time scale $t_{\eta}$ is 
\begin{equation}
    t_{\eta} = \frac{\eta^2}{\nu}.
    \label{eq:dissipation-time-NEP}
\end{equation}
As Table~\ref{table:turbulence-NEP} shows, we find $t_{\eta}=49\pm15$ kyr. Finally, combining the dissipation scale and the dissipation time, the energy transfer rate $\epsilon$ is
\begin{equation}
    \epsilon = \frac{\eta^2}{t_{\eta}^3}.
    \label{eq:energy-transfer-rate-NEP}
\end{equation}
We find $\epsilon=(1.3\pm0.5)\,\times 10^{-4}$\,L$_{\odot}$M$_{\odot}^{-1}$. This is comparable to the value obtained by \cite{hennebelle_turbulent_2012} in the CNM ($\epsilon_{\rm CNM} \sim$ 10$^{-3}$\,L$_{\odot}$M$_{\odot}^{-1}$). As noted by these authors, $\epsilon_{\rm CNM}$ is considerably lower than the energy transferred to the ISM via UV and visible radiation from stars. It is therefore also the case for the WNM. 

We now consider the case where ambipolar diffusion due to ion-neutral friction is the dominant mechanism of energy dissipation. 
Following \cite{miville-deschenes_structure_2017}, the ambipolar diffusion typical scale is 
\begin{equation}
    l_{\rm AD} = \sqrt{\frac{\pi}{\mu m_{\rm H}}} \frac{B}{2X \left< \sigma v \right> n^{3/2}}
    \label{eq:ambipolar-diffusion-scale}
\end{equation}
where $X=n_e/(n_e+n)$ is the ionization ratio with an electron density $n_e=0.0213$\,cm$^{-3}$ \citep{berkhuijsen_filling_2006}, $\left< \sigma v \right>=2\times10^{-9}$\,cm$^3$\,s$^{-1}$ is the collision rate between ions and neutral assumed to be the Langevin rate, and $B=6\,\mu$G is the typical value of the magnetic field strength in the WIM/WNM of the solar neighborhood \citep{beck_galactic_2001}. We find $l_{\rm AD}=(3.3\pm0.7)\,\times10^{-2}$\,pc. This is typically a factor 3-4 higher than the dissipation scale inferred from the molecular viscosity. Given the possible variations of ionization fraction or magnetic field strength, these two scales could be comparable. Therefore, it seems difficult to conclude which mechanism dominates the turbulent energy dissipation in the WNM of NEP. 

\section{Thermal instability} \label{sec:thermal-instability}

It is believed that the condensation mode of TI is responsible for the thermal condensation of the warm neutral phase of the ISM that leads to the formation of CNM structures in galaxies. Building on previous Sect.~\ref{sec:gas-properties}, the thermal and turbulent properties of the WNM in NEP allows us to evaluate if perturbations around the mean thermodynamic state of the gas allows the condensation mode of TI to develop and then to grow freely. Values calculated in this section are tabulated in Table~\ref{table:turbulence-NEP}.

\subsection{Development of the condensation mode} \label{subsec:condensation-mode}

\begin{figure}[!t]
 \centering
 \includegraphics[width=\linewidth]{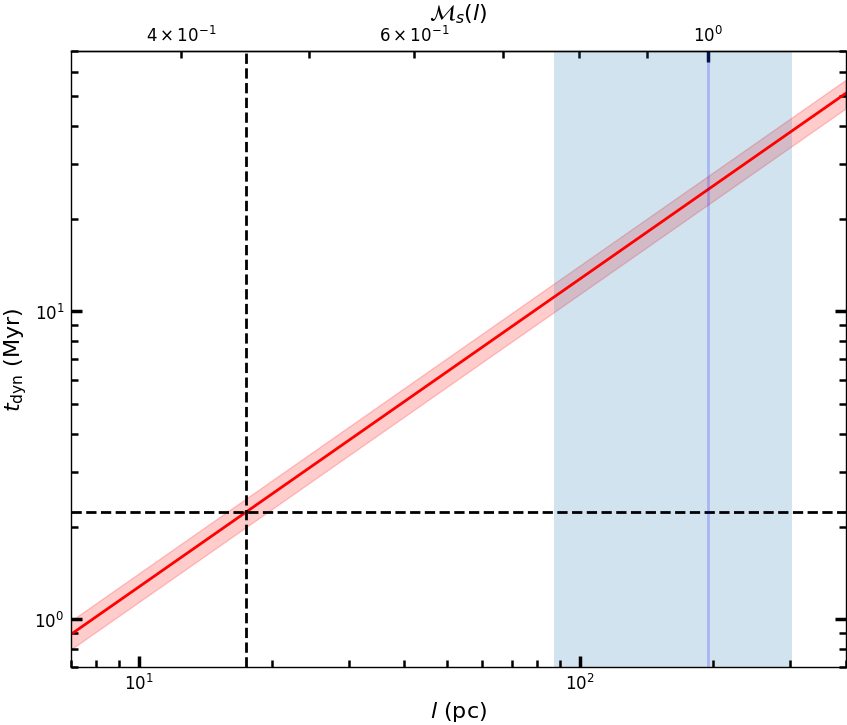}
 \caption{Dynamical time vs scale (red). Top axis gives the corresponding turbulent sonic Mach number. Blue line and shade area shows the sonic scale and its error. Vertical and horizontal black dashed lines show the cooling length and cooling time, respectively.}
 \label{fig:tdyn_tcool}
\end{figure}

In a idealized non-viscous static fluid in thermal equilibrium, the isobaric criterion for development of the condensation mode of thermal instability can be expressed as
\begin{equation}
    \left(\frac{\partial P}{\partial n}\right)_{\mathcal{L}=0} < 0
    \label{eq:criterion-condensation}
\end{equation}
\citep{field_thermal_1965,wolfire_neutral_1995}. One might expect the WNM to be in a thermal pressure and volume density range which, if slightly compressed, can fall into the range where Eq.~\ref{eq:criterion-condensation} is satisfied. As Fig.~\ref{fig:wolfire_NEP_cross} shows, the thermodynamical state of the WNM in NEP is compatible with the development of the condensation mode of TI.

\subsection{Can the condensation mode grow freely~? Cooling and dynamical time scales} \label{subsec:dynamical}
When TI develops, a condensation is possible if the cooling time of the fluid element is shorter than its dynamical time\footnote{Otherwise, the energy lost by radiation is small when compared to the increase of internal energy and the process can be considered as adiabatic (i.e., no transfer of heat between the fluid element and its surrounding medium)} \citep{hennebelle_dynamical_1999}
\begin{equation}
    \mathcal{I}(l) = \frac{t_{\rm cool}}{t_{\rm dyn}(l)} < 1,
    \label{eq:condensation-criterion-NEP}
\end{equation}
where the cooling time is 
\begin{equation}
    t_{\rm cool} = \frac{k_B T_k}{(\gamma - 1) n \Lambda(T_k)}.
\end{equation}
with $\Lambda(T_k)$ the net cooling function. The dynamical time of the compression is
\begin{equation}
     t_{\rm dyn}(l) = \frac{l_{\rm pc}}{C_s},
    \label{eq:dynamical-time-NEP}
\end{equation}
where $l_{\rm pc}$ is the linear size of the homogeneous compression.

In a turbulent fluid, compression events arise on a continuous range of scales characterized by the energy cascade.
Figure~\ref{fig:tdyn_tcool} shows the dynamical time as function of scale (see top axis for the corresponding turbulent sonic Mach number) that depends linearly on the sound speed of the WNM. The horizontal black dashed line shows the cooling time, $t_{\rm cool}\sim2.3$\,Myr, based on the standard model ($N_c=1\times10^{20}$\,cm$^{-2}$ and $G_0=1.7$) from \cite{wolfire_neutral_2003}. Above this line, $\mathcal{I}$ is satisfied, ensuring that TI can grow freely. The intersection between $t_{\rm dyn}(l)$ and $t_{\rm cool}$ corresponds to the linear scale that satisfies $\mathcal{I}=1$ and which is called the cooling length $\lambda_{\rm cool}\simeq18$\,pc\footnote{The cooling length is also the scale at which the WNM is non-linearly unstable \citep{audit_thermal_2005}.}. In other words, compression arising on linear scales $l_{\rm pc} > \lambda_{\rm cool}$ also satisfies Eq.~\ref{eq:condensation-criterion-NEP}.
By construction, we find that $\mathcal{M}_s(\lambda_{\rm cool})=0.45$, showing that a too weak turbulent flow prevent TI to grow freely to enable the formation of CNM structures, in accordance with numerical experiment of bi-stable turbulent flows \citep{saury_structure_2014}.

For comparison, the blue line and shade area shows the sonic scale and its error $\lambda_s=(1.7\pm1.1)\,\times10^2$\,pc, that satisfies $\mathcal{M}_s(\lambda_s)=1$. At this scale, the information travels at sound speed $C_s$ and $t_{\rm dyn}(\lambda_s)=25\pm12$\,Myr. We find $\mathcal{I}(\lambda_s)=0.09\pm0.04$.

\subsection{The Field Length} \label{subsec:field-length}
During this condensation phase, along with heating and cooling, the fluid is subjected to thermal diffusion. The length at which radiative heating and cooling processes become comparable to thermal diffusion is the Field length\footnote{The Field length is also called the conduction length and can be seen as the typical size of the front between the WNM and the CNM.}
\begin{equation}
    \lambda_{F} = \sqrt{ \frac{\kappa(T_k) \, T_k}{n^2 \, \Lambda(T_k)} }\,,
    \label{eq:field-length-NEP}
\end{equation}
\citep{field_thermal_1965,begelman_global_1990} where $\kappa = 2.5 \times$ 10$^{3}T_k^{1/2}$\,erg\,cm$^{-1}$\,K$^{-1}$\,s$^{-1}$ is the thermal conductivity for hydrogen atoms \citep{parker_instability_1953}. We find $\lambda_{F}=0.10\pm0.06$\,pc. $\lambda_{F}$ is considerably lower than $\lambda_{\rm cool}$, ensuring that compression of linear size $l_{\rm pc}$ > $\lambda_{\rm cool}$ will be more affected by cooling than thermal diffusion.

\section{Summary}
\label{sec:summary}

We have presented a study of the thermal and turbulent properties of the multiphase neutral ISM in the solar neighbourhood based on the analysis of 21\,cm emission data of a high Galactic latitude field (NEP) from the GHIGLS \HI\ survey \citep{martin_ghigls:_2015}. In particular we have presented an original method to extract information from a 21\,cm emission data cube allowing us to constrain the physical properties of the WNM. 

The specificity of this study relies on the following aspects:
\begin{itemize}
    \itemsep0em
    \item The Gaussian decomposition tool \texttt{ROHSA} \citep{marchal_rohsa:_2019} was used to provide a spatially coherent model of the local phases: WNM, LNM and CNM. This allowed us to produce maps of the WNM column density and centroid velocity.
    
    \item The distance and physical scales of the \HI\ gas observed in projection were determined using 3D dust extinction map.
    \item The thermal and turbulent contribution to the WNM line widths were disentangled using the observed properties of its centroid velocity field and CNM structures within it.
\end{itemize}

Being able to isolate the WNM from the total \HI\ emission cube, to locate where it is coming from along the line-of-sight, and to estimate the contribution of turbulent motions to the line width, allowed us to estimate the physical properties of the WNM in great detail.
The main conclusions are as follows.
\begin{itemize}
    \itemsep0em
    \item The WNM average density ($n=0.74\pm0.41$\,cm$^{-3}$), kinetic temperature ($T_k=6.0\pm1.3\times 10^3$\,K) and thermal pressure ($P_{th}/k_{B}=4.4\pm 2.6\times10^3$\,K\,cm$^{-3}$) are compatible with expected values for the Solar neighborhood conditions assuming typical cooling and heating processes.
  
    \item The mass fraction of each phase show significant variations over the $12^\circ \times 12^\circ$ field. 
    On average the phase proportions in the local velocity component are CNM 8\%, LNM 28\%, and WNM 64\%. We estimated the volume filling fraction of the WNM to be $f_\mathcal{V}^{\rm WNM}=0.6\pm0.2$.
    
    \item The WNM has the properties of a trans-sonic turbulent flow, with low density contrast ($\sigma_{\rho/\rho_0} = 0.6\pm0.2$), density and velocity power spectra compatible with $k^{-11/3}$, and a low Mach number at the largest scale probed ($\mathcal{M}_s(l=130\,{\rm pc}) = 0.87\pm0.15$). The WNM turbulent velocity dispersion at a scale of 1\,pc ($\sigma_{v_z}(1) = 0.76 \pm 0.1$\,km\,s$^{-1}$) is similar to what is observed in denser phases of the ISM. 
    
    \item Our determinations of the dynamic and static scales involved in the thermal condensation of the WNM confirm that the relatively low strength of turbulence coupled to the thermal state of the gas allow the condensation mode of TI to grow freely to form cold dense structures observed in the CNM. 
    \item We confirm the theoretical prediction of the cooling length $\lambda_{\rm cool}\sim18$\,pc \citep{hennebelle_dynamical_1999} in the WNM of the solar neighborhood. In addition we put an observational constrain on the Field length, also coherent with theoretical expectations.
\end{itemize}

\acknowledgments

This work was supported by the Natural Sciences and Engineering Research Council of Canada (NSERC).
Part of this work was supported by Hyperstars, a project funded by the MASTODONS initiative of the CNRS mission for inter-disciplinarity. This work took part under the program Milky-Way-Gaia of the PSI2 project funded by the IDEX Paris-Saclay, ANR-11-IDEX-0003-02. 
We gratefully acknowledge Peter G. Martin and Benjamin Godard for enlightening conversations, Claire Murray for providing us the result of the 21-Sponge survey analysis, and Ludovic Montier for providing us an Healpix version of the 3D dust map used in this work. We also gratefully acknowledge John Dickey and Robert Benjamin for reviewing this work as members of A.M. thesis jury. We thank the anonymous referee whose comments and suggestions have improved this manuscript.

\vspace{5mm}
\software{matplotlib \citep{hunter_2007}, NumPy \citep{van_der_walt_2011}, and Astropy\footnote{\url{http://www.astropy.org}}, a community-developed core Python package for Astronomy \citep{astropy_2013, astropy_2018}.}

\appendix

\section{Power spectrum analysis}
\label{app:sps-procedure}

Following \cite{martin_ghigls:_2015}, the power spectrum $P(k)$ of $C(\rb)$ (eqv. $\nh(\rb)$) is the azimuthal average of the modulus of its Fourier transform, and is modelled as 
\begin{equation}
    P(k) = B(k) \times P_0 k^{\gamma} + N(k)\,,
    \label{eq:model-Pk}
\end{equation}
where $P_0$ is the amplitude of the power spectrum, $\gamma$ is the scaling exponent, $B(k)$ is the beam of the instrument assuming a 2D Gaussian of FWHM = 9\farcm 4, and $N(k)$ is the noise estimated by taking the power spectrum of empty channels of the PPV cube. Finally, edges of the image are apodized using a cosine function to avoid effects due to the replication done by the Fourier transform algorithm.

\section{Impact of volume filling factor on the statistics of projected quantities}
\label{app:volume-filling-Pk}

The interpretation of power law slopes of projected quantities, like the column density and the centroid velocity, has been the subject of several theoretical and numerical studies in the past 20 years \citep[e.g.,][]{miville-deschenes_use_2003}. The general consensus is that the power spectrum of column density, estimated from optically thin lines, is a reliable proxy for the power spectrum of the 3D density field. As mentioned earlier, the centroid velocity is a reliable proxy for the velocity field only in the limit of small 3D density contrasts. 

All the previous studies devoted to the understanding of the link between the statistical properties of projected and 3D fields are based on the assumption that the fluid that projects on the sky fills the 3D volume completely. In reality this is rarely satisfied. For instance, the properties of molecular clouds are often analyzed using line tracers like $^{12}$CO or $^{13}$CO that depend strongly on the local gas volume density. This provides information only about the densest part of the fluid in 3D. 

One could think that \HI\ is less affected by this density effect and that the 21\,cm emission would be a more reliable tracer of the whole 3D volume. This is most probably the case for the whole 21\,cm emission but then the difficulty lies in the estimate of the statistical properties of a 3D multi-phase fluid with density contrasts of the order of 100-1000. 

The separation of the thermal phases from the 21\,cm emission data with {\tt ROHSA} allows the extraction of the low density WNM, opening the possibility to study interstellar turbulence by comparing its multi-scale statistics of density and velocity to the ones of controlled numerical experiments of isothermal turbulence. 
In this context, one interesting question is to what extent the fact that WNM does not occupy the full volume on the line of sight affects the relationship between the statistical properties of the 3D fields (density and velocity) and the projected quantities (column density and centroid velocity). To our knowledge, this has never been explored specifically. 

The evaluation of the effect of a partial filling of a fluid on the projected quantities is clearly beyond the scope of this paper. In fact it is a rather ill-defined problem as one needs to define the statistics of the physical fields (density, velocity) as well as the 3D shape of the volume occupied by the fluid. That shape of that mask is likely to be defined by a density threshold in the case of CO emission or CNM structures. For the more diffuse inter-cloud \HI\ medium (LNM and WNM), the shape of the volume is probably related to the temperature field, linked to heating and cooling processes. So it could be partly related to the density field but also to the radiation field intensity field. A proper study of this effect should be done with dedicated numerical simulations of the multi-phase ISM. In order to explore the main trends of the expected effects, we present a short study based only on fractional Brownian motion fields.

\subsection{Density field}
\begin{figure}[!t]
  \centering
  \includegraphics[width=0.5\linewidth]{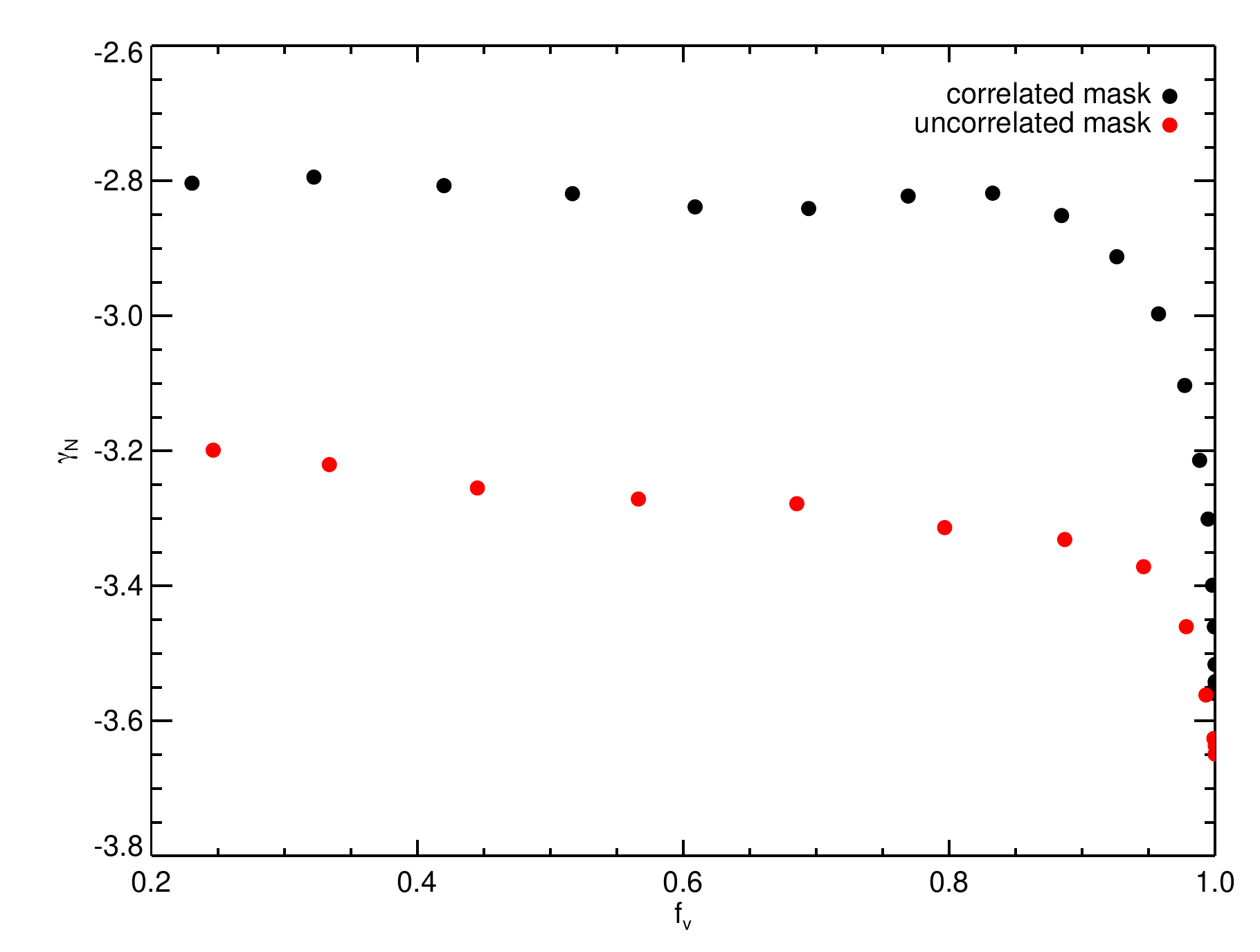}
  \caption{Power spectrum slope of the column density resulting from the projection of 3D density fields with partial 3D sampling.}
  \label{fig:gamma_N_vs_fv}
\end{figure}

Let's first look at the effect on the column density. To do so we have constructed 3D density fields $n_{\rm H}(\rb)$ over a grid of $128\times128\times256$ pixels. We have chosen to produce 3D cubes with a depth ($L_z$) twice the size in $[x,y]$ to mimic an observation like NEP. 
The statistics of $n_{\rm H}(\rb)$ is assumed to follow a log-normal distribution, with $\delta(n_{\rm H})/<n_{\rm H}>=1$, and a 3D power spectrum slope of $\gamma_{n}=-11/3$, reminiscent of what is seen for compressible, sub or trans-sonic turbulent flows. 

With such low value of $\delta(n)/<n>$, the power spectrum of the column density $N_{\rm H} = \sum_z n_{\rm H}\, dz$ integrated over the full volume has the same slope as the 3D density field : $\gamma_N = \gamma_n$. The question is what is the power spectrum slope of $N_{\rm H}$ if a fraction of the volume is removed from $n_{\rm H}$. We have explored two possibilities: 1) a threshold in density and 2) a mask independent of density. In the first case we have gradually put to zero the density values above a given threshold. As the threshold is lowered, the volume filling factor of the gas gets smaller. 

In the second case, $n_{\rm H}(\rb)$ is put to zero in region based on a second independent field $m(\rb)$. At this point it is difficult to evaluate what should be the statistics of $m(\rb)$. We decided to use a positive, low contrast ($\sigma(m)/<m>$ = 1/5) Gaussian field with a power spectrum slope of -11/3. Like for the first case, region of $n_{\rm H}$ are put to zero but here we use the criteria on $m$, where $m > m_{0}$. As $m_{0}$ is lowered, larger and larger region of $n_{\rm H}$ are put to zero, lowering the volume filling factor. 
 
Figure~\ref{fig:gamma_N_vs_fv} shows the power spectrum slope of $\gamma_N$ recovered in both cases, as a function of the volume filling factor $f_v$. The black and red points corresponds to case 1) and 2) respectively. As expected, when the fluid occupies the full volume ($f_v=1$), we recover the result $\gamma_N = \gamma_n$. The main result we obtained is that, as the volume filling factor is lowered, the power spectrum slope increases (the $P(k)$ is flatter). Interestingly it seems that $\gamma_N$ reaches an almost constant value for $f_v < 0.8$, but that value depends on the way the 3D mask is built. 

The main result here is the fact that a volume filling factor lower than unity introduces a systematic bias in the determination of the 3D density slope from the column density. A value of $f_v < 1$ systematically flattens the power spectrum. The power spectrum slope $\gamma_N$ obtained from observations provides an upper limit of the true value of $\gamma_n$. The effect is important. According to our little experiment it could be as much as $\gamma_n = \gamma_N-1$. This has important consequences on the use of the column density power spectrum slope to constrain the properties of turbulence, the Mach number in particular \citep{kim_density_2005}.

\begin{figure}[!t]
  \centering
  \includegraphics[width=0.5\linewidth]{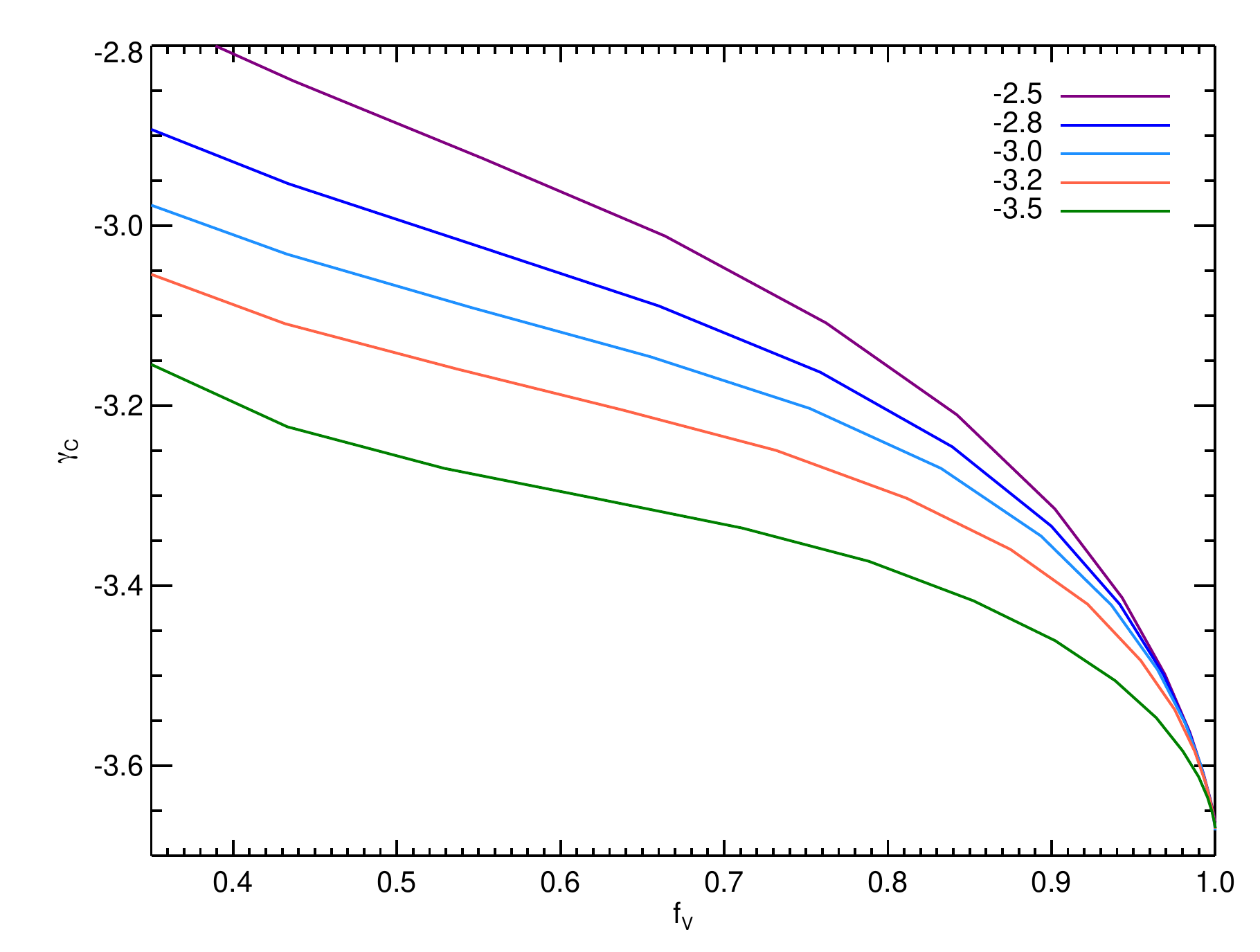}
  \caption{Power spectrum slope of the centroid velocity field as a function of the volume filling factor of the fluid. The masked regions are defined using a gradual threshold on a log-normal density field with a power spectrum slope $\gamma_n$. The different curves show the result for a range of $\gamma_n$.}
  \label{fig:gamma_C_vs_fv}
\end{figure}

\subsection{Velocity field}

We developed a similar experiment for the velocity field. We built uncorrelated 3D density and velocity cubes, $n(\rb)$ and $v_z(\rb)$, using the same method as previously. The velocity field has a Kolmogorov slope ($\gamma_v=-11/3$). The masking is done on $n$, by removing pixels in 3D where the density if above a given threshold $n>n(0)$. Then the centroid velocity field $C(\rb)$ is estimated by calculating the average velocity on each line of sight, using only the unmasked pixels. 
Figure~\ref{fig:gamma_C_vs_fv} shows the power spectrum slope of $\gamma_C$ recovered in each case, as a function of the volume filling factor $f_v$.
Like for the column density field case, the partial filling of the 3D volume introduces a bias in the projected velocity field. We note though that the effect is less dramatic for the velocity centroid than for the column density field.

\bibliographystyle{aasjournal}
\bibliography{myref}{}


\end{document}